\documentclass[fleqn,usenatbib]{mnras}

\usepackage{newtxtext,newtxmath}

\usepackage[T1]{fontenc}
\usepackage{ae,aecompl}
\usepackage{color}
\usepackage{xspace}
\usepackage{subcaption}
\usepackage{graphicx}	
\usepackage{amsmath}	
\usepackage{amssymb}	
\usepackage{longtable}
\usepackage{multicol}
\usepackage{adjustbox}
\usepackage{lipsum}     
\usepackage{scalerel}
\usepackage{tikz}
\usetikzlibrary{svg.path}

\definecolor{orcidlogocol}{HTML}{A6CE39}
\tikzset{
  orcidlogo/.pic={
    \fill[orcidlogocol] svg{M256,128c0,70.7-57.3,128-128,128C57.3,256,0,198.7,0,128C0,57.3,57.3,0,128    ,0C198.7,0,256,57.3,256,128z};
    \fill[white] svg{M86.3,186.2H70.9V79.1h15.4v48.4V186.2z}
                 svg{M108.9,79.1h41.6c39.6,0,57,28.3,57,53.6c0,27.5-21.5,53.6-56.8,53.6h-41.8V79.1z M    124.3,172.4h24.5c34.9,0,42.9-26.5,42.9-39.7c0-21.5-13.7-39.7-43.7-39.7h-23.7V172.4z}
                 svg{M88.7,56.8c0,5.5-4.5,10.1-10.1,10.1c-5.6,0-10.1-4.6-10.1-10.1c0-5.6,4.5-10.1,10.    1-10.1C84.2,46.7,88.7,51.3,88.7,56.8z};
  }
}

\newcommand\orcidicon[1]{\href{https://orcid.org/#1}{\mbox{\scalerel*{
\begin{tikzpicture}[yscale=-1.,transform shape]
\pic{orcidlogo};
\end{tikzpicture}
}{|}}}}
\usepackage{hyperref}   

\newcommand{\shade}{SH$\alpha$DE\xspace} 

\newcommand{\mstar}{$M_\star$\xspace}
\newcommand{\halpha}{$\mathrm{H\alpha}$\xspace}

\newcommand{\kms}{\,km\,s$^{-1}$}
\newcommand{\HI}{H{\small I}}
\newcommand{\HII}{H{\small II}}

\defcitealias{barat+2019}{B19}
\defcitealias{Federrath2017}{F17}
\defcitealias{delosReyes+Kennicutt2019}{RK19}

\title[\shade\ Survey]{\texorpdfstring{\boldmath\shade}{Lg}: Survey description and mass--kinematics scaling relations for dwarf galaxies}

\author[Barat et al.]{Dilyar Barat$^{1,2}$\orcidicon{0000-0002-4972-8188}\thanks{E-mail: Dilyar.Barat@anu.edu.au},
Francesco D'Eugenio$^{3}$\orcidicon{0000-0003-2388-8172},
Matthew Colless$^{1,2}$\orcidicon{0000-0001-9552-8075},
Sarah M. Sweet$^{2,4}$\orcidicon{0000-0002-1576-2505},\newauthor
Brent Groves$^{1,2,5}$\orcidicon{0000-0002-9768-0246} and 
Luca Cortese$^{2,5}$\orcidicon{0000-0002-2959-1014}\\
$^{1}$Research School of Astronomy and Astrophysics, Australian National University, Canberra, ACT 2611, Australia\\
$^{2}$ARC Centre of Excellence for All Sky Astrophysics in 3 Dimensions (ASTRO 3D), Australia\\
$^{3}$Sterrenkundig Observatorium, Universiteit Gent, Krijgslaan 281 S9, B-9000 Gent, Belgium\\
$^{4}$School of Mathematics and Physics, University of Queensland, Brisbane, QLD 4072, Australia\\
$^{5}$International Centre for Radio Astronomy Research (ICRAR), University of Western Australia, Crawley, WA 6009, Australia
}

\date{Accepted 2020 September 3. Received 2020 August 14; in original form 2020 June 15}

\pubyear{2020}

\begin{document}
\label{firstpage}
\pagerange{\pageref{firstpage}--\pageref{lastpage}}
\maketitle

\begin{abstract}
The Study of \halpha from Dwarf Emissions (\shade) is a high spectral resolution (R=13500) \halpha integral field survey of 69 dwarf galaxies with stellar masses $10^6<M_\star<10^9 M_\odot$. The survey used FLAMES on the ESO Very Large Telescope. \shade\ is designed to study the kinematics and stellar populations of dwarf galaxies using consistent methods applied to massive galaxies and at matching level of detail, connecting these mass ranges in an unbiased way. In this paper we set out the science goals of \shade, describe the sample properties, outline the data reduction and analysis processes. We investigate the $\log{M_{\star}}-\log{S_{0.5}}$ mass--kinematics scaling relation, which have previously shown potential for combining galaxies of all morphologies in a single scaling relation. We extend the scaling relation from massive galaxies to dwarf galaxies, demonstrating this relation is linear down to a stellar mass of $M_{\star}\sim10^{8.6}M_{\odot}$. Below this limit, the kinematics of galaxies inside one effective radius appear to be dominated by the internal velocity dispersion limit of the \halpha-emitting gas, giving a bend in the $\log{M_{\star}}-\log{S_{0.5}}$ relation. Replacing stellar mass with total baryonic mass using gas mass estimate reduces the severity but does not remove the linearity limit of the scaling relation. An extrapolation to estimate the galaxies' dark matter halo masses, yields a $\log{M_{h}}-\log{S_{0.5}}$ scaling relation that is free of any bend, has reduced curvature over the whole mass range, and brings galaxies of all masses and morphologies onto the virial relation.
\end{abstract}



\section{Introduction}

Dwarf galaxies are the most common galaxies in the Universe. In the current paradigm of galaxy formation, they are the building blocks of larger galaxies, so understanding their properties is key to understanding the cosmic process of structure formation \citep[e.g.][]{searle+zinn1978}. Thanks to large numerical simulations \citetext{e.g. EAGLE, \citealp{schaye+2015}; HorizonAGN, \citealp{dubois+2014}; IllustrisTNG, \citealp{springel+2018}; and Romulus, \citealp{tremmel+2017}}, the spatial distribution of cosmic structures is understood relatively well \citetext{e.g. \citealp{artale+2017}, but see \citealp{hatfield+2019} for a different view}. However, these simulations have insufficient resolution to accurately simulate galaxies with stellar masses $M_\star < 10^9 \, \mathrm{M_\odot}$, and so cannot reliably predict the properties of dwarf galaxies. In addition, these large cosmological simulations include a number of simplifications, so-called `subgrid  physics', which implement the effect of physical processes on scales that are currently impossible to simulate. These include star formation \citep[e.g.][]{dutton+2019}, stellar feedback and supernova feedback \citep[e.g.][]{hopkins+2014, marinacci+2019}, active-galactic nuclei feedback \citep[AGN; e.g.][]{booth+schaye2009}, metal diffusion in and outside the inter-stellar medium \citep[e.g.][]{hafen+2019}, but also purely gravitational collisions \citetext{so-called softening; see e.g. \citealp{vogelsberger+2019}, their Table 2}.

The impact of subgrid physics (and of its implementation) changes with galaxy properties. Given that \mstar is one of the most fundamental galaxy properties, dwarf galaxies represent an invaluable testbed, because they allow us to study how the effect of these different physical processes change below $M_\star = 10^9 \, \mathrm{M_\odot}$, the dwarf-galaxy mass threshold adopted in this work. In fact, while regular galaxies span already three orders of magnitude in stellar mass ($10^9 < M_\star < 10^{12}\, \mathrm{M_\odot}$), including dwarf galaxies doubles the baseline in \mstar, adding the mass range between $10^6$ and $10^9 \, \mathrm{M_\odot}$.

Despite the desirability of including dwarf galaxies in large extragalactic surveys, the study of these low-mass systems is hampered by their defining physical properties: dwarf galaxies are less luminous than regular galaxies, so observations require longer integration times and/or larger collecting areas; they are smaller, so studying their structure requires better spatial resolution; finally, dwarf galaxies have lower velocity dispersions, so an unbiased measurement of their kinematics requires either high spectral resolution or very high signal to noise \citep[cf. ][their sec.\ 2.2.2]{zhou+2017}.

In light of these obstacles, it is not surprising that dwarf galaxies have been mostly left out of the the integral field spectroscopy revolution of this decade: for example, SAMI has a mass limit of $10^8 \, \mathrm{M_\odot}$ \citep{croom+2012, bryant+2015} while MaNGA has a mass limit of $10^9 \, \mathrm{M_\odot}$ \citep{bundy+2015}. This gap is partially filled by studies of local dwarf galaxies, but these works do not employ the same methods as large extragalactic surveys: they rely either on neutral hydrogen observations \citep[e.g.\ ][]{hunter+2012}, or on individually-resolved stars \citep[e.g.\ ][]{tolstoy+2009}, neither of which is yet available beyond the local Universe. A notable exception is represented by the SIGRID survey \citep{nicholls+2011}, however even the high spectral resolution of SIGRID ($R=7000$) is insufficient to probe the regime of thermal broadening ($\lesssim 15 \, \mathrm{km \, s^{-1}}$) that might bias dynamical scaling relations of dwarf galaxies. In summary, at present, no optical survey can simultaneously deliver sufficient numbers, spatial resolution, and spectral resolution to reliably study the kinematics and dynamics of dwarf galaxies, and to compare them to more massive galaxies.

The \shade\ survey was designed to fill this gap, and to deliver a sample of 69 dwarf galaxies with high spectral and spatial resolution \halpha observations. The survey was designed with four goals in mind: (i)~to test the linearity of galaxy scaling relations over a range in mass and with sufficient spectral resolution to not be observationally limited; (ii)~to measure and explain $f_\mathrm{asym}$, the fraction of dwarf galaxies with asymmetric kinematics; (iii)~to study the dynamical effect of star-formation feedback in the low-mass regime; and (iv)~to study angular momentum accretion.

The main goals of this paper are to introduce the \shade\ survey and to present our results for dwarf scaling relations. The paper is organised as follows: in Section~\ref{s.g} we outline the scientific goals of \shade; Section~\ref{s.d} presents the selection criteria and sample, the observations, and the data reductions; the analysis of this data is presented in Section~\ref{sec:dataanalysis}. We then focus on the first of the survey goals,  providing our analysis of mass--kinematics scaling relations for the \shade\ galaxies in Section~\ref{s.res}; in Section~\ref{s.dis} we discuss our results; finally, Section~\ref{s.sum} provides a summary of our findings.

Throughout this paper we assume a $\Lambda$CDM cosmology with $\Omega_M$=0.3, $\Omega_{\lambda}$=0.7 and $H_0$=70\kms, and a Chabrier initial mass function \citep{chabrier2003}. All magnitudes are in the AB system \citep{oke+gunn1983}.

\section{Goals of  \texorpdfstring{\shade}{Lg}}\label{s.g}

Dwarf galaxies with stellar mass $M_{\star} < 10^{9}\,\mathrm{M_\odot}$ are special compared to normal galaxies with  $10^{9} < M_{\star} < 10^{12}\,\mathrm{M_\odot}$. The low masses of dwarf galaxies, and the environments in which they reside, make them interesting targets which can challenge theories of galaxy formation and evolution that are based on massive galaxies. This section outlines the types of experiments that can be carried out using the \shade\ observations, including galaxy scaling relations, kinematic asymmetries, star formation and ISM turbulence, and angular momentum accretion in dwarf galaxies.  

\subsection{Galaxy scaling relations}\label{s.g.sr}

For disc galaxies, optical luminosity correlates with \HI\ 21\,cm line width; this is the Tully-Fisher (TF) relation \citep{TullyFisher1977}. Since the discovery of this scaling relation, a plethora of studies have been carried out investigating the scaling relation across multiple photometric bands \citep[for a summary see][]{Ponomareva+2017}. The TF relation has been widely used in determining galaxy distances, and subsequently measuring cosmological constants and galaxy flows \citep[e.g.][]{Courtois+Tully2012}. The TF relation is also an important tool in testing various theories of gravity \citep[e.g.][]{Milgrom1983,Sanders+1990, Mo+1998, McGaugh2012, Desmond+2015}.

The stellar mass TF relation (where luminosity is replaced by stellar mass) has long been observed to have a `knee' at low circular velocity where the slope of the relation steepens \citep{matthews+1998, McGaugh+2000, Amorin+2009, McGaugh+2010, Sales+2017}. This region of the relation is predominantly occupied by dwarf galaxies, which are found to have on average larger gas fractions than regular star-forming galaxies \citep{McGaugh+2000, hunter+2012, lelli+2014, oh+2015}. Further \HI\ follow-up of dwarf galaxies showed that, by including the cold gas mass and so using the total baryonic mass instead of stellar mass, a linear TF relation is restored over 5 orders of magnitude in mass \citep{McGaugh+2000, McGaugh2005b, Lelli+2016, Iorio+2017}. This illustrates both the importance of extending observations of galaxy scaling relations to dwarf galaxies and of understanding the physical basis of such relations.

At the massive-galaxy end of the TF relation, it has been shown that even passive, early-type 
galaxies obey the TF relation whenever they include enough gas to obtain reliable measurements of the circular velocity \citep{denheijer+2015}. However, early-type galaxies differ from late-type galaxies in that most early-type galaxies do not have detectable \HI\ gas \citep{Cortese2020}, and their kinematics are dominated by unordered (or at least complex) motions, observed as velocity dispersion, rather than the highly ordered motions observed as rotation velocity. The Faber-Jackson (FJ) relation \citep{FaberJackson1976} for early-type galaxies is the correlation of their velocity dispersions with their luminosities or stellar masses. 

Tightly-correlated TF or FJ relations require reliable morphological selection of (respectively) late-type or early-type galaxy samples, which is time-consuming and difficult. The desirability of unifying these relations in a generalized kinematic scaling relation that applies to galaxies of all types, which became possible with the advent of integral field spectroscopy (IFS), led to the construction of the $S_{0.5}$ parameter \citep{weiner+2006} defined as:
\begin{equation}
    S_{0.5} \equiv \sqrt{0.5 V_\mathrm{rot}^2 + \langle \sigma_\mathrm{los} \rangle^2}
\end{equation}
where $V_\mathrm{rot}$ is the rotation velocity and $\langle \sigma_\mathrm{los} \rangle$ is the average line-of-sight velocity dispersion. This parameter can be thought as a proxy for the circular velocity, with a uniform asymmetric drift correction, independent of galaxy morphology. Despite the simplistic approximation, $S_{0.5}$ correlates tightly with stellar mass, and---crucially---this correlation holds for all morphological types and for kinematics measured either from stars or warm ionised gas \citep{cortese+2014}. The $\log M_{\star} - \log S_{0.5}$ relation being, in both these senses, more universal than the TF or FJ relations, can be a powerful probe of galaxy dynamics \citep[e.g.][]{oh+2016, Cannon+2016}, structure formation \citep[e.g.][]{Dutton+2012, Tapia+2017, Desmond+2019}, and can be used to measure distances and peculiar velocities and so cosmological parameters \citep[e.g.][]{McGaugh2012,  glazebrook2013, Said+2015}. In fact, galaxy formation theory in the context of the $\Lambda$CDM model predicts that the baryon fraction $f_b$ increases with halo mass $M_h$, and peaks at $M_h \approx 10^{12}\,\mathrm{M_\odot}$ \citep[e.g.][]{moster+2013}. Observations appear to confirm this expectation: within one effective radius ($R_e$), regular galaxies are baryon dominated \citep[e.g.][]{cappellari+2013a}, whereas dwarf galaxies seem to be dark matter dominated at all radii \citep{penny+2009}. If this scenario is correct, we expect this to be reflected in the $\log M_{\star} - \log S_{0.5}$ relation: when the dynamics become dominated by dark matter, stellar or baryonic mass will be a less precise tracer of the dynamics, due to the relatively large scatter in the $f_b$--$M_h$ relation \citep[e.g.][]{dicintio+lelli2015, desmond2017}. Alternatively, dwarf galaxies might be baryon-dominated within one $R_e$ \citep{sweet+2016}, and claims of dark-matter-dominated dwarfs may stem from non-equilibrium dynamics in tidal dwarf galaxies.

The $\log M_{\star} - \log S_{0.5}$ relation is linear over three orders of magnitude in mass \citep[e.g.][]{cortese+2014}, but, within limits imposed by the current mass range and spectral resolution, it {\em appears} to become steeper below $M_\star = 10^9\,\mathrm{M_\odot}$ \citep[][hereafter: \citetalias{barat+2019}]{barat+2019}. This value is intriguingly close to the theoretical predictions \citetext{\citealp{cortese+2014}, \citealp{Aquino-Ortiz2018}, \citealp{Gilhuly+2019}}; on the other hand, the fact that break in the $\log M_{\star} - \log S_{0.5}$ relation occurs at different values of \mstar for the stellar and gas kinematics, and just below the instrument spectral resolution in each case, suggests that measurement systematics might also play a role, producing inflated velocity dispersions that make the relation appear steeper \citepalias{barat+2019}.

In summary, based on current observational evidence is still unclear whether the change of slope in the $\log M_{\star} - \log S_{0.5}$ relation is due to increasing gas fractions \citep[as for the TF relation, cf.][]{McGaugh+2000}, insufficient spectral resolution \citepalias[cf.][]{barat+2019}, non-equilibrium dynamics \citep[a natural consequence of the hypothesis in ][]{sweet+2016}, or increasing dark matter fraction below $M_\star \approx 10^{8}$--$10^{9}\,\mathrm{M_\odot}$ \citep[predicted by e.g.][]{behroozi+2013}. Part of the uncertainty is due to the fact that current IFS surveys are designed to probe galaxies with $M_\star \gtrsim 10^9\,\mathrm{M_\odot}$, so that we lack accurate data precisely where the relation becomes most interesting. It is clear that obtaining new data with better spectral resolution will extend the baseline in \mstar and better constrain the $\log M_{\star} - \log S_{0.5}$ relation.

\subsection{Kinematic asymmetries in dwarf galaxies}

The hypothesis that the bend in the $\log M_{\star} - \log S_{0.5}$ relation is due to non-equilibrium dynamics of the warm ionised gas tracer is plausible. Because of Malmquist bias, magnitude-limited samples of star-forming dwarf galaxies may have higher-than-average SFR per unit mass (specific SFR; sSFR), which is associated with mergers \citep{robaina+2009} and/or substantial accretion of cold gas \citep{elmegreen+2012, thorp+2019}. Indeed, \citet{bloom+2017a} find that a fraction $f_\mathrm{asym}\sim0.5$ of isolated dwarf galaxies exhibit irregular gas kinematics, inconsistent with a rotating disc. These disturbances increase the scatter at the low-mass end of the TF relation \citep{bloom+2017b}, and therefore might also contribute to steepening the low-mass end slope of the $\log M_{\star} - \log S_{0.5}$ relation. Whereas asymmetric gas kinematics in massive galaxies is usually explained by recent galaxy-galaxy interactions, this is not the case for dwarf galaxies---these low-mass systems differ in two fundamental aspects from their high-mass counterparts. Firstly, asymmetric kinematics: dwarfs are found predominantly in isolation, thus ruling out a dominant role for tidal interactions. Secondly, most dwarfs have regular photometric shapes that are inconsistent with recent substantial mergers.

It is possible that these galaxies are accreting relatively large amounts of unseen neutral hydrogen from the intergalactic medium, because their halo mass is smaller than the quenching threshold \citep[cf.][]{elmegreen+2012}. In this case, we expect $f_\mathrm{asym}$ to be insensitive to \mstar for dwarf galaxies. Alternatively, \citet{bloom+2018} propose that the asymmetries are caused by the discrete distribution of giant molecular clouds, which becomes more coarse with decreasing stellar mass. In this case, $f_\mathrm{asym}$ must strongly increase with decreasing \mstar. Probing the gas kinematics well below $M_\star = 10^9 \, \mathrm{M_\odot}$ could discriminate between these two models.

\subsection{The link between star formation and ISM turbulence}

Star-forming dwarf galaxies have higher SFR per unit mass on average than regular galaxies. This fact follows from the sub-linear slope of the star-forming sequence, $\log \mathrm{SFR} = (0.67\pm0.08) \log M_\star$ (\citealp{noeske+2007}; see also \citealp{speagle+2014}, \citealp{renzini+peng2015}), so that the sSFR decreases with \mstar. With their high sSFR, star-forming dwarf galaxies can also be used to probe the interplay between star formation and gas dynamics. Part of the \halpha velocity dispersion is due to turbulent motions, which are thought to regulate the conversion of gas into stars and are therefore key to understanding galaxy formation and evolution \citep[e.g.][]{Green+2010, federrath+klessen2012, padoan+2014}. However, the origin of this turbulence is not well understood: possible mechanisms include star formation feedback, inter-cloud collisions \citep{tasker+tan2009}, gas accretion \citep{klessen+hennebelle2010}, galactic shear within the gas disc \citep{Krumholz+Burkhart2016}, and magnetorotational instability \citep{tamburro+2009}. Recent IFS surveys have clarified that the relation between gas velocity dispersion and SFR originates from a local relation \citep{Lehnert+2009, Lehnert+2013}. \citet{zhou+2017} found that the observed random motions of the star-forming gas require additional sources beyond star-formation feedback. However, gas turbulence is of the order of 10--15\,$\mathrm{km \, s^{-1}}$, smaller than the spectral resolution of all large IFU surveys, which may therefore introduce a systematic bias in the measurements. To test for such bias, a sample of local star-forming galaxies observed at high spectral resolution could help constrain theoretical models.

\subsection{Angular momentum accretion of dwarf galaxies}

The relation between the fundamental parameters of stellar mass \mstar and angular momentum $J_\star$ has been empirically studied since \citet{Fall1983}, who found that $j_\star=q M_\star^\alpha$, where $j_\star = J_\star/M_\star$ is the specific stellar angular momentum. The scale factor $q$ varies with morphology in the sense that bulge-dominated galaxies have a lower angular momentum than disk-dominated galaxies of the same mass. The exponent $\alpha \sim 2/3$ agrees with the analogous relation between halo specific angular momentum $j_h$ and halo mass $M_h$ in a scale-free cold dark matter universe, namely $j_h \propto M_h^{2/3}$ \citep{Mo+1998}. In addition, the mean specific angular momentum of the baryons $j_b$ is within a factor of two of the halo $j_h$ \citep{Barnes+1987,Catelan+1996a,Catelan+1996b,Posti+2018MN}. While the broad connection between $j_b$ and $j_h$ is still a topic of active research \citep[e.g.][]{Jiang+2019,Posti+2019}, the $M_\star$--$j_\star$ relation is generally assumed to be a byproduct of hierarchical assembly, since galaxy mergers increase both mass and angular momentum for the haloes as well as for their stars \citep{Lagos+2018}.

Low-mass galaxies are thought to be the building blocks of more massive galaxies, both in mass and angular momentum. However, while the $M_\star$--$j_\star$ relation has been confirmed over a broad range of morphologies \citep{RF12,OG14,Cortese+2016,FR18,Sweet+2018,Posti+2018AA}, these works \citep[with the exception of][]{Posti+2018AA}, have focussed on more massive galaxies with stellar mass $M_\star \gtrsim 10^9\,\mathrm{M_\odot}$. Moreover, dwarf galaxies are fundamentally different to massive galaxies: the interrelated properties of morphology, gas fraction, star formation rate, and metallicity are not related to stellar mass by a simple power-law \citep[e.g.][]{Scodeggio+2002,Dekel+2003,Salim+2007,Tremonti+2004}. Neither does it follow that the relation between $M_\star$ and $j_\star$ should take the form of an unbroken power law over large dynamic range in $M_\star$. Indeed, the semi-analytic models Dark SAGE \citep{Stevens+2016} and GALFORM \citep{Mitchell+2018}, cosmological zoom-in simulations NIHAO \citep{Obreja+2016}, and cosmological hydrodynamical simulation EAGLE \citep{Lagos+2017} all find that simulated disk-dominated low-mass ($M_\star \lesssim 10^{9.5}\,\mathrm{M_\odot}$) galaxies have more angular momentum than predicted by a constant $\alpha \sim 2/3$ slope. This increased angular momentum could be related to an enhanced gas fraction and/or lower velocity dispersion for dwarf galaxies \citep{Obreschkow+2016}. A hint of such an elevation above $\alpha = 2/3$ is seen in \citet{FR13}, which includes just four galaxies with masses in the range $10^{8.5} < M_\star < 10^{9.5}\,\mathrm{M_\odot}$. \citet{Posti+2018AA} extended the $M_\star$--$j_\star$ relation to stellar masses as low as $M_\star \gtrsim 10^7\,\mathrm{M_\odot}$ and found a constant $\alpha = 0.55$ for all stellar masses, which is at odds with the predictions of $\alpha \sim 2/3$ for massive galaxies and a shallower slope at dwarf masses. However, while \citet{Posti+2018AA} measured the relation down to $M_\star \gtrsim 10^7\,\mathrm{M_\odot}$, the sample only included six galaxies with $10^7 < M_\star < 10^8\,\mathrm{M_\odot}$. The lowest-mass galaxies have very small uncertainties, but those with $10^8 < M_\star < 10^{9.5}\,\mathrm{M_\odot}$ seem to show a $j_\star$ above the fitted relation, in line with \citet{FR13} and simulations such as \citet{Stevens+2016}. To test the robustness of this previous finding, and determine whether or not dwarf galaxies have elevated angular momentum,  a larger sample extending to yet smaller stellar masses is needed.

\section{Data}\label{s.d}

 The \shade\ sample consists of 49 star-forming galaxies selected form the Sloan Digital Sky Survey Data Release 12 \citep[hereafter: SDSS;][]{eisenstein+2011, alam+2015}, as well as 20 targets from the SAMI survey as control sample. These targets are observed with high spectral resolution IFU FLAMES instrument, and reduced with the standard data reduction package. Other ancillary data are obtained from the SDSS DR12. Details of these processes are outlined in the subsections below. 

\begin{figure}
  \centering
  \includegraphics[type=pdf,ext=.pdf,read=.pdf,width=1.0\columnwidth]{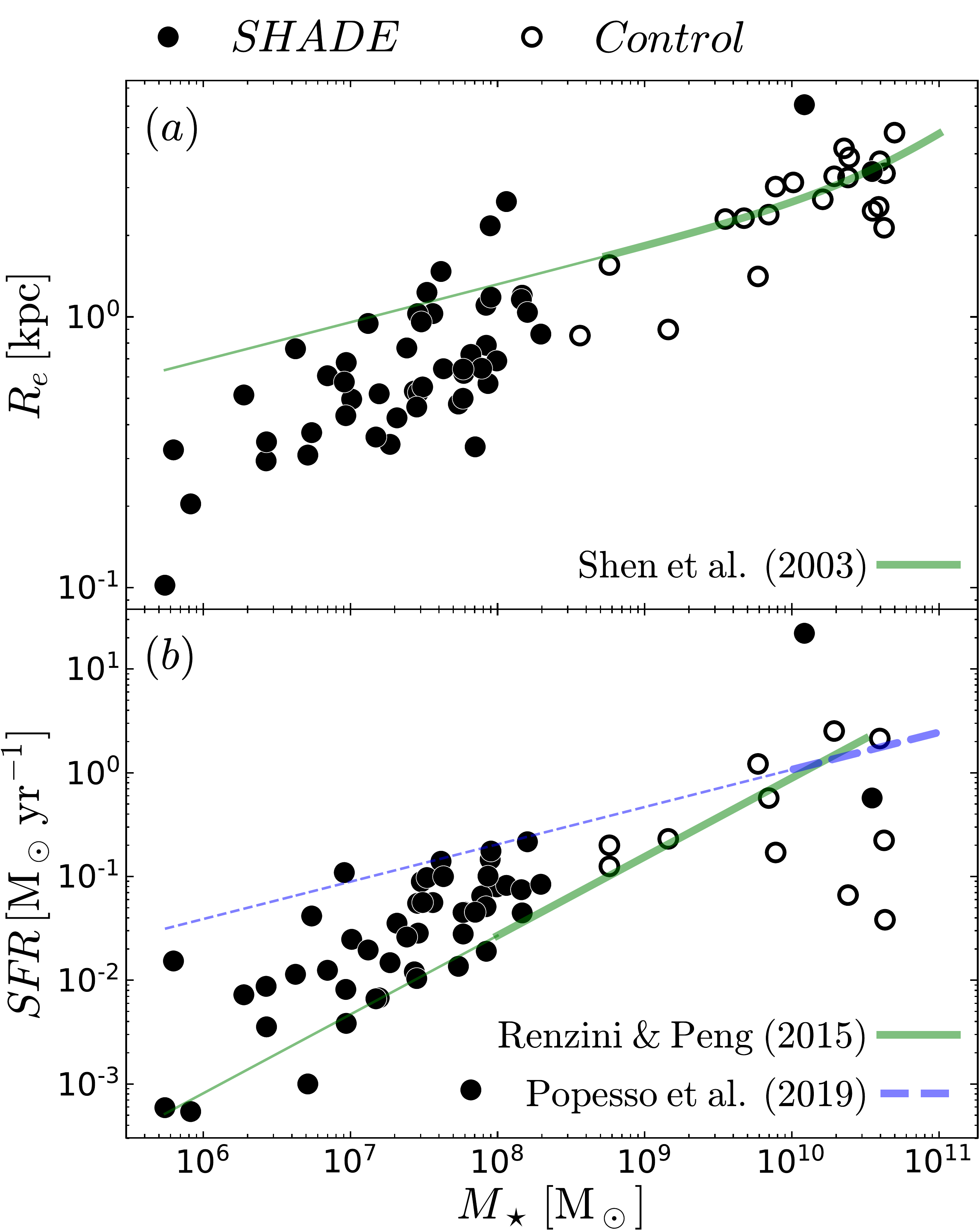}
  {\phantomsubcaption\label{f.sad.sample.a}
   \phantomsubcaption\label{f.sad.sample.b}}
  \caption{
  The \shade\ galaxies (filled circles) lie close to both the mass--size relation (panel~\subref{f.sad.sample.a}) and the star-forming sequence (panel~\subref{f.sad.sample.b}) of local star-forming galaxies. The \shade\ Survey also includes a control sample of 20 non-dwarf galaxies, drawn from the SAMI Survey (open circles). The lines are best-fit relations drawn from the literature (thich/thin lines mark the fitted/extrapolated domain of each relation).}\label{f.sad.sample}
\end{figure}

\subsection{The Sample}\label{s.d.ss.sample}

The sample was designed to probe, as uniformly as possible, the low-mass end of the galaxy distribution. We selected our targets from SDSS by applying four constraints: (i)~stellar masses in the range $10^5 \leq  M_\star  \leq 10^{8.5} \mathrm{M_\odot}$; (ii)~apparent sizes in the range $1.2 \leq R_d \, \mathrm{[arcsec]} \leq 11$; (iii)~\halpha fluxes above a threshold $f_\mathrm{H\alpha} > 5 \times 10^{-16} \mathrm{[erg \, s^{-1} \, cm^{-2} \, \text{\AA}^{-1} \, arcsec^{-2}]}$; and (iv)~targets observable in the relevant semester (March to August). The first three criteria ensure (i)~coverage of the mass range relevant to our goals; (ii)~sufficient spatial resolution, while fitting within the instrument field of view (FOV); and (iii)~sufficient signal-to-noise ratio (SNR) to measure \halpha kinematics. These criteria and the spectral resolution of the instrument also mean we have sufficient spectral/velocity resolution (even at the low end of the mass range) to measure the expected velocity dispersions based on extrapolating the $\log M_{\star} - \log S_{0.5}$ relation.

We visually inspected this set of 601 galaxies and rejected 75 objects that proved to be artefacts or galaxies with significant contamination from interlopers or neighbours (including mergers). For each target galaxy the FLAMES instrument requires a guide star, which we selected from the GAIA Survey \citep{gaia+2016}; as a result, we removed another 104 valid targets that had no suitable guide star in GAIA public data release 2 \citep{gaia+2018}. From this final pool of 422 galaxies, we scheduled 49 targets for observation using a custom scheduler that aims at uniformly sampling the target mass range while ensuring that no target had airmass larger than 1.5 at any point during the observation.


\begin{table}
  \begin{center}
  \setlength{\tabcolsep}{2pt}
  \caption{Summary of observations: the main sample.}
  \label{t.sad.obs.main}
  \begin{tabular}{crrrrrr}
  \hline
  \multicolumn{1}{c}{SH$\alpha$DE} & \multicolumn{1}{c}{SDSS} &  \multicolumn{1}{c}{$f_\mathrm{H\alpha}$} & \multicolumn{1}{c}{$T_\mathrm{exp}$} & \multicolumn{1}{c}{$\langle \rm{AM} \rangle$} & \multicolumn{1}{c}{Seeing} & \multicolumn{1}{c}{Run} \\
    \multicolumn{1}{c}{ID} & \multicolumn{1}{c}{SpecObjID}  & \multicolumn{1}{c}{$\dagger$} & \multicolumn{1}{c}{$s$} & & \multicolumn{1}{c}{$\mathrm{arcsec}$} & \\
  \multicolumn{1}{c}{(1)} & \multicolumn{1}{c}{(2)} & \multicolumn{1}{c}{(3)} & \multicolumn{1}{c}{(4)} & \multicolumn{1}{c}{(5)} & \multicolumn{1}{c}{(6)} & \multicolumn{1}{c}{(7)}  \\
  \hline
1	&	344558384307005440	&	8.17$\pm$1.59	&	$900 \times 3$	&	1.12	&	0.92	&	A \\
2	&	385144108173256704	&	7.66$\pm$3.31	&	$900 \times 3$	&	1.12	&	0.83	&	A \\
9	&	388552594202585088	&	7.10$\pm$4.25	&	$900 \times 3$	&	1.96	&	1.22	&	B \\
42	&	585558149897938944	&	8.72$\pm$0.63	&	$900 \times 3$	&	1.28	&	0.99	&	A \\
46	&	585553751851427840	&	9.50$\pm$0.64	&	$900 \times 3$	&	1.12	&	1.33	&	A \\
55	&	666581432214251520	&	15.96$\pm$0.73	&	$500 \times 2$	&	1.57	&	1.34	&	B \\
56	&	1326366762505103360	&	10.35$\pm$2.07	&	$500 \times 2$	&	1.27	&	0.6	    &	A \\
59	&	736372110183655424	&	9.49$\pm$0.13	&	$900 \times 3$	&	1.03	&	0.89	&	B \\
64	&	743256427883161600	&	21.76$\pm$0.92	&	$500 \times 2$	&	1.16	&	1.16	&	B \\
128	&	1025746258003781632	&	10.30$\pm$2.01	&	$500 \times 2$	&	1.17	&	1.29	&	A \\
132	&	1033705622308153344	&	10.39$\pm$0.61	&	$500 \times 2$	&	1.15	&	1.66	&	A \\
136	&	667815084277393408	&	25.33$\pm$3.56	&	$500 \times 2$	&	1.16	&	0.83	&	A \\
137	&	1159814941646022656	&	9.23$\pm$3.31	&	$900 \times 3$	&	1.17	&	0.72	&	B \\
148	&	1151799775146829824	&	6.32$\pm$0.33	&	$900 \times 3$	&	1.16	&	1.17	&	A \\
151	&	1105748373979293696	&	7.29$\pm$1.16	&	$900 \times 3$	&	1.14	&	0.75	&	A \\
152	&	1150797565365610496	&	6.56$\pm$4.37	&	$900 \times 3$	&	1.24	&	0.81	&	A \\
165	&	459448556403582976	&	16.12$\pm$0.69	&	$500 \times 2$	&	1.12	&	0.73	&	B \\
171	&	1128315028525574144	&	11.57$\pm$0.72	&	$500 \times 2$	&	1.26	&	1.54	&	A \\
194	&	1819558756229343232	&	6.70$\pm$2.57	&	$900 \times 3$	&	1.73	&	1.04	&	A \\
200	&	1375985247325284352	&	11.72$\pm$0.98	&	$500 \times 2$	&	1.22	&	0.6	    &	A \\
218	&	1816204695102842880	&	6.63$\pm$1.33	&	$900 \times 3$	&	1.28	&	0.77	&	A \\
231	&	1939939348900243456	&	30.66$\pm$4.00	&	$500 \times 2$	&	1.95	&	0.64	&	B \\
260	&	1829747929997404160	&	5.57$\pm$0.47	&	$900 \times 3$	&	1.2	    &	0.74	&	A \\
271	&	2014355951074699264	&	6.97$\pm$2.58	&	$900 \times 3$	&	1.23	&	0.88	&	A \\
283	&	1946713989006256128	&	10.24$\pm$0.71	&	$500 \times 2$	&	1.3	    &	0.8	    &	B \\
284	&	2058148387030591488	&	46.53$\pm$1.42	&	$500 \times 2$	&	1.18	&	0.65	&	A \\
286	&	3321458402864949248	&	7.41$\pm$3.64	&	$900 \times 3$	&	1.25	&	1.26	&	A \\
288	&	2049141199587010560	&	10.15$\pm$1.66	&	$500 \times 2$	&	1.17	&	0.65	&	A \\
311	&	2485035976443848704	&	8.58$\pm$2.03	&	$900 \times 3$	&	1.42	&	0.99	&	A \\
314	&	1234065512632182784	&	21.22$\pm$11.44	&	$500 \times 2$	&	1.3	    &	0.69	&	A \\
315	&	431222718478706688	&	10.91$\pm$0.74	&	$500 \times 2$	&	1.3	    &	0.85	&	A \\
319	&	1150711803458643968	&	5.82$\pm$0.25	&	$900 \times 3$	&	1.1	    &	1.3	    &	A \\
320	&	1155193413423884288	&	7.15$\pm$0.36	&	$900 \times 3$	&	1.1	    &	1.45	&	A \\
322	&	1254417201423738880	&	33.39$\pm$11.03	&	$500 \times 2$	&	1.64	&	0.77	&	A \\
323	&	1157472431619729408	&	8.58$\pm$4.37	&	$900 \times 3$	&	1.26	&	0.69	&	A \\
327	&	1166516459447281664	&	5.80$\pm$2.05	&	$900 \times 3$	&	1.14	&	0.7	    &	B \\
330	&	425751828999727104	&	6.01$\pm$0.47	&	$900 \times 3$	&	1.42	&	0.84	&	A \\
331	&	1150794541708634112	&	6.53$\pm$2.98	&	$900 \times 3$	&	1.29	&	1.16	&	A \\
343	&	1219400493801957376	&	8.16$\pm$6.26	&	$900 \times 2$	&	1.52	&	1.52	&	A \\
344	&	1676573588242589696	&	6.64$\pm$0.40	&	$900 \times 3$	&	1.13	&	0.86	&	B \\
352	&	1214947748281870336	&	5.02$\pm$0.56	&	$900 \times 3$	&	1.14	&	0.67	&	B \\
424	&	3087280562621147136	&	17.63$\pm$3.15	&	$500 \times 2$	&	1.32	&	1.47	&	A \\
430	&	3088358084452575232	&	6.52$\pm$0.45	&	$900 \times 3$	&	1.37	&	1.38	&	A \\
431	&	3099604713595758592	&	7.38$\pm$0.46	&	$900 \times 3$	&	1.27	&	2.01	&	A \\
433	&	3091856729999173632	&	17.83$\pm$5.95	&	$500 \times 2$	&	1.87	&	0.95	&	B \\
435	&	3132266537138808832	&	7.57$\pm$0.55	&	$900 \times 3$	&	1.43	&	1.35	&	A \\
469	&	2931991316961191936	&	11.77$\pm$0.75	&	$500 \times 2$	&	1.37	&	0.92	&	A \\
496	&	952674913926277120	&	12.22$\pm$0.79	&	$500 \times 2$	&	1.17	&	0.81	&	A \\
520	&	1156448780808120320	&	10.71$\pm$2.31	&	$500 \times 2$	&	1.51	&	0.87	&	A \\
  \end{tabular}
  \end{center}$\dagger$ Units of $10^{-16} \, \mathrm{erg \, s^{-1} \,
  cm^{-2} \, \text{\AA}^{-1} \, arcsec^{-2}}$. Source: SDSS DR12 \citep{Thomas+2011}
  
\end{table}

\begin{table}
  
  \begin{center}
  
  \setlength{\tabcolsep}{2pt}
  
  \caption{Summary of observations: the control sample, drawn from SAMI.}
  \label{t.sad.obs.control}
  \hskip-1.4cm
  \begin{tabular}{crrrrrrrr}
  \hline
  \multicolumn{1}{c}{\shade / GAMA} & \multicolumn{1}{c}{$f_\mathrm{H\alpha}$} & \multicolumn{1}{c}{$T_\mathrm{exp}$} & \multicolumn{1}{c}{$\langle \rm{AM} \rangle$} & \multicolumn{1}{c}{Seeing} & \multicolumn{1}{c}{Run} \\
    \multicolumn{1}{c}{ID}  & \multicolumn{1}{c}{$\dagger$} & \multicolumn{1}{c}{$s$} & & \multicolumn{1}{c}{$\mathrm{arcsec}$} & \\
  \multicolumn{1}{c}{(1)} & \multicolumn{1}{c}{(2)} & \multicolumn{1}{c}{(3)} & \multicolumn{1}{c}{(4)} & \multicolumn{1}{c}{(5)} & \multicolumn{1}{c}{(6)}  \\
  \hline
      106049 &  4.78$\pm$1.94 & $900 \times 3$ & 1.53 & 1.06 & B \\
      296934 &  28.65$\pm$7.44& $900 \times 3$ & 1.29 & 0.71 & C \\
      319150 &  4.77$\pm$1.74 & $900 \times 3$ & 1.29 & 1.12 & B \\
      511921 &  13.92$\pm$3.47& $900 \times 3$ & 1.75 & 0.95 & B \\
      594906 &  30.35$\pm$3.60& $900 \times 3$ & 1.96 & 1.05 & B \\
  9008500333 &  2.75$\pm$0.93 & $900 \times 3$ & 1.10 & 0.79 & B \\
  9008500356 &  1.50$\pm$0.63 & $900 \times 3$ & 1.10 & 1.32 & B \\
  9011900125 &  -           & $900 \times 3$ & 1.11 & 0.72 & B \\
  9011900128 &  -           & $900 \times 3$ & 1.39 & 0.83 & B \\
  9016800065 &  2.13$\pm$0.51& $900 \times 3$ & 1.12 & 0.68 & B \\
  9016800314 &  1.36$\pm$0.40& $900 \times 3$ & 1.12 & 1.33 & B \\
  9091700123 &  -           & $900 \times 3$ & 1.35 & 0.89 & B \\
  9091700137 &  -           & $900 \times 3$ & 1.02 & 1.13 & B \\
  9091700444 &  -           & $900 \times 3$ & 1.11 & 1.20 & B \\
  9239900178 &  3.02$\pm$1.15& $900 \times 3$ & 1.75 & 1.31 & B \\
  9239900182 &  19.10$\pm$1.88& $900 \times 3$ & 1.05 & 1.38 & B \\
  9239900237 &  16.28$\pm$2.62& $900 \times 3$ & 1.23 & 1.11 & B \\
  9239900246 &  -           & $900 \times 3$ & 1.35 & 1.05 & B \\
  9239900370 &  -           & $900 \times 3$ & 1.20 & 0.92 & B \\
  9388000269 &  -           & $900 \times 3$ & 1.06 & 1.15 & B \\
  \end{tabular}
  \end{center}
  $\dagger$ Units of $10^{-16} \, \mathrm{erg \, s^{-1} \,
  cm^{-2} \, \text{\AA}^{-1} \, arcsec^{-2}}$. Not used for the control sample.
\end{table}

The main \shade\ sample is complemented with 20 targets from the SAMI Survey to be used as a control sample; these additional galaxies were randomly selected within $10^{8.5}<M_{\star}<10^{10.5} \rm M_{\odot}$ to span a broader range of masses, to ensure a consistency in the $S_{0.5}$ measurements across SAMI and SHADE. 

Note that 2 non-control high-mass ($M_{\star}>10^{10} \rm M_{\odot}$) galaxies are present in the \shade\ sample, this is because their stellar mass values from SDSS used for target selection were significantly lower in comparison to the method used in this paper. Details on our stellar mass measurement is described in section \ref{subsec:ancillarydata}.  

The position of the \shade\ and control galaxies on the mass--size plane is shown in Figure~\ref{f.sad.sample.a} (filled and empty circles, respectively). Both sets of galaxies have sizes that are not inconsistent with the local mass--size relation for star-forming galaxies \citep[solid green line; ][]{shen+2003}. The mass--size relation is measured only down to $10^{8.5}\,\mathrm{M_\odot}$: below this lower limit we simply extrapolated the best-fit function (indicated by the thin section of the line). The \shade\ galaxies lie systematically below this extrapolation in radius, but are consistent with a single linear mass--size relation spanning the entire mass range.  For the star-forming sequence, the \shade\ galaxies lie systematically above the local relation and its extrapolation to lower masses \citep[][green line in Figure~\ref{f.sad.sample.b}]{renzini+peng2015}, as expected given that we selected bright \halpha emitters (cf.\ Section~\ref{s.d.ss.sample}). However, other authors have reported a shallower slope for the star-forming sequence \citetext{e.g. \citealp{popesso+2019}; dashed blue line in Figure~\ref{f.sad.sample.b}}; the \shade\ galaxies lie systematically below a (more extreme) extrapolation of this relation.

\subsection{Observations}\label{s.d.ss.obs}

We present new data from the FLAMES instrument at the $8\,\mathrm{m}$ VLT Unit Telescope~2 \citep{pasquini+2002}, using the ARGUS IFU  and the GIRAFFE optical spectrograph. ARGUS is a rectangular array of $22\times14$ square microlenses; at the 1:1 scale, each microlens samples 0.52\,arcsec and the IFU FOV spans $11.4\times7.3$\,arcsec$^2$. Light from individual microlenses in the IFU is fed to GIRAFFE using optical fibres. Besides the IFU itself, ARGUS also provides 15 dedicated sky fibres, which can be placed anywhere inside the instrument FOV. Within the slit, the fibres are arranged in 15 bundles, each consisting of 20 object fibres and a single sky fibre. On the detector, the fibres are separated by 5.3\,pixels, whereas the bundles are separated by 12--20\,pixels.

We used setup L682.2, consisting of the low-resolution grating centred at $\lambda_\mathrm{central} = 6822$\,\AA\ and the LR06 filter. This configuration delivers a nominal spectral resolution $R = 13500$ ($\Delta\lambda = 0.505$\,\AA) and the spectrum is sampled with 0.2\,\AA\ pixels. The wavelength range is 6440--7160\,\AA, covering the rest-frame \halpha emission at 6562.8\,\AA\ up to redshift $z \approx 0.08$, appropriate for our target selection. The detector read speed was set to 50\,kHz and high gain, because at this spectral resolution our data is limited by read-out noise.

The observations (see Tables~\ref{t.sad.obs.main} \&~\ref{t.sad.obs.control}) were carried out under ESO program 0101.B-0505 in two Visitor Mode runs (A and B), complemented by a Service Mode run (C) allocated as time compensation. Each target was observed for either 17\,min (for targets brighter than $f_\mathrm{H\alpha} > 5 \times 10^{-16} \, \mathrm{erg \, s^{-1} \, cm^{-2} \, \text{\AA}^{-1} \, arcsec^{-2}}$) or 45\,min; the integrations were split into either two 8.5\,min-long exposures or three 15\,min-long exposures. The median airmass-corrected seeing was 0.88\,arcsec, with a dispersion of 0.32\,arcsec (values are full-width at half maximum, FWHM).

\subsection{Data reduction}\label{s.d.ss.dr}

\begin{figure}
  \centering
  \includegraphics[type=pdf,ext=.pdf,read=.pdf,width=\columnwidth]{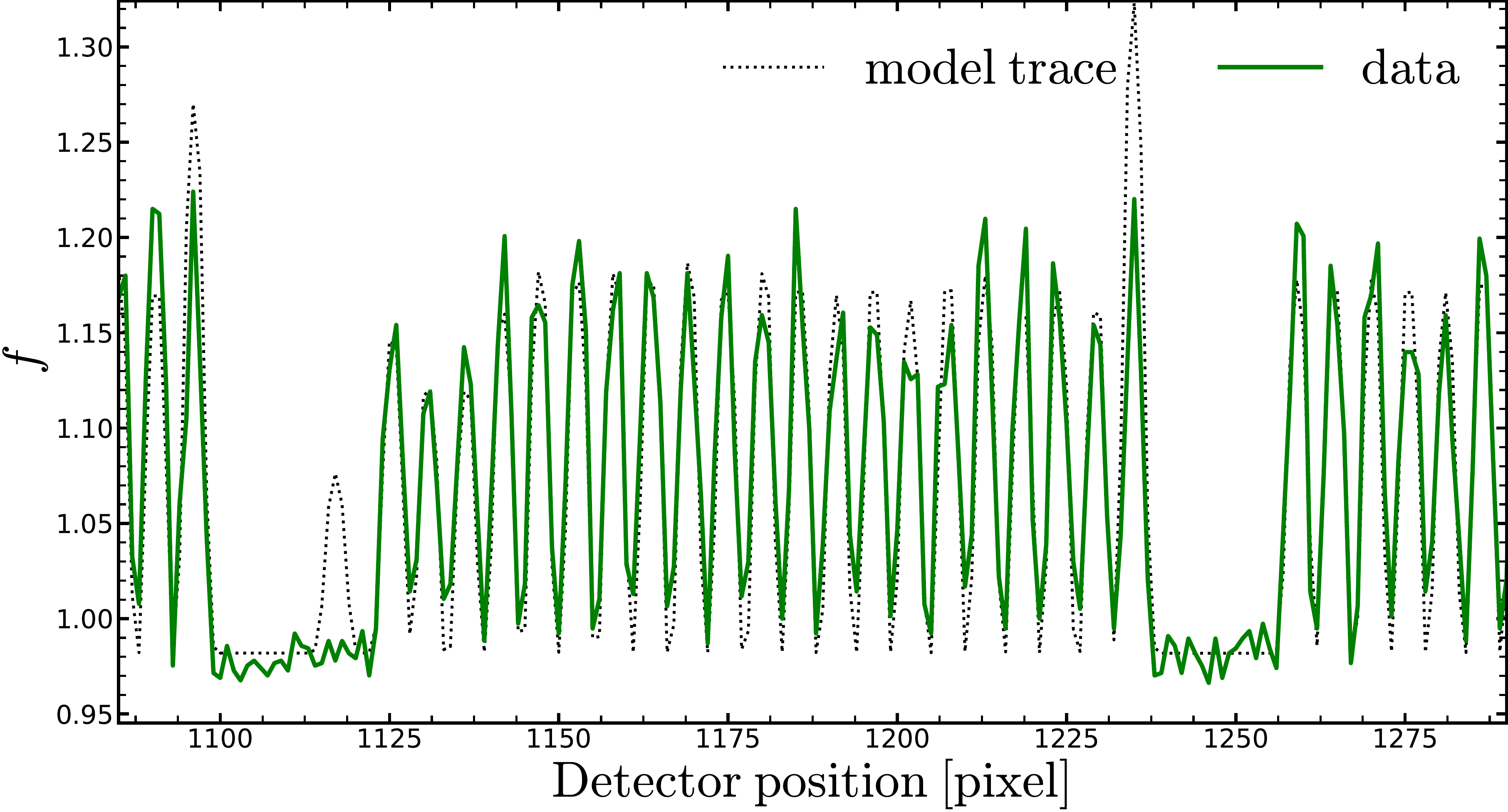}
  \caption{Comparison of a section of the model for the spatial distribution of light derived from a flat-field frame with data from a science frame. The model trace (dotted black line) is the multiple Gaussian fit to a flat-field frame; the science trace (solid green line) is the region around the sky line at $\lambda = 6864.97$\,\AA\ for a science frame. For display purposes, we show only one bundle, consisting of the 21 fibre traces between the two troughs; the fibre present in the fibre-flat model but absent in the science frame is used for simultaneous wavelength calibration and was switched off in our setup.}\label{f.sad.traces}
\end{figure}

We perform a standard data reduction\footnote{The current reduced data is not flux-calibrated, but flux-calibration will be performed for the official data release (Sweet et~al. in~prep.).} using the \texttt{giraf-kit} package provided by ESO \citep{blecha+2000, royer+2002}. The bias level was estimated using the overscan regions. For each galaxy, we traced the centroid and width of the spectra using the most recent Nasmyth flat, and derived both the flat field and the scattered light model. For each row along the dispersion direction, we sum the flux inside the fibre trace \cite[optimal extraction,][is not yet implemented]{horne1986}; given that the science spectra are too faint to trace their position on the CCD, we use the fibre traces derived from the Nasmyth flat-field frames. This approach is justified by a direct comparison of the flat-field traces to the science traces, which can be determined around bright emission sky lines (Figure~\ref{f.sad.traces}).

\subsubsection{Wavelength calibration}\label{s.d.ss.dr.sss.wavecal}

The wavelength calibration relies on dedicated Th-Ar lamp exposures. In the relevant function of \texttt{giraf-kit}, the dispersion solution consists of an optical model (which predicts the position of each line on the detector), plus a polynomial correction \citep{royer+2002}. The free parameters are constrained iteratively using the position of 70 unsaturated emission lines on the detector. We validate the resulting calibration using prominent emission sky lines taken directly from the science frames. First we created a model continuum spectrum, consisting of the sum of the galaxy and sky continua. We smoothed the spectra with a median filter (kernel width of 10.2\;\AA, or 51\,pixels), we masked the regions within five FWHM from any emission line, and we fitted a spline to the smooth, masked spectrum. This model was then subtracted from the observed spectra to obtain an emission-line spectrum.

We selected a list of 17 bright singlet lines from the UVES Atlas \cite[][Table~\ref{t.sad.skylines}]{hanuschik2003} and fitted each line with a Gaussian, allowing for uniform background\footnote{We tested the use of doublet emission lines by fitting the lines simultaneously, but found the measured line widths have larger scatter than the singlet emission lines and so discarded doublets.}. We used the bounded least squares algorithm from \texttt{scipy.optimize}, which in turn relies on the Trust Region Reflective algorithm for minimisation \citep{branch+1999}. As initial values, we used the intensity, central wavelength, and intrinsic line dispersion reported in \citet{hanuschik2003}.

As a diagnostic of the wavelength calibration, we take the relative offset $\lambda_m/\lambda_t-1$ between the measured line wavelength and the value tabulated in the UVES Atlas. We find that both the precision and the accuracy of the wavelength solution are excellent: the standard deviation of the relative offset is $2.6 \times 10^{-6}$ (0.017\,\AA\ in absolute terms), the mean offset is $(-2.5\pm0.6) \times 10^{-6}$ ($-0.016\pm0.004$\,\AA).

\begin{figure*}
  \centering
  \includegraphics[type=pdf,ext=.pdf,read=.pdf,width=\textwidth]{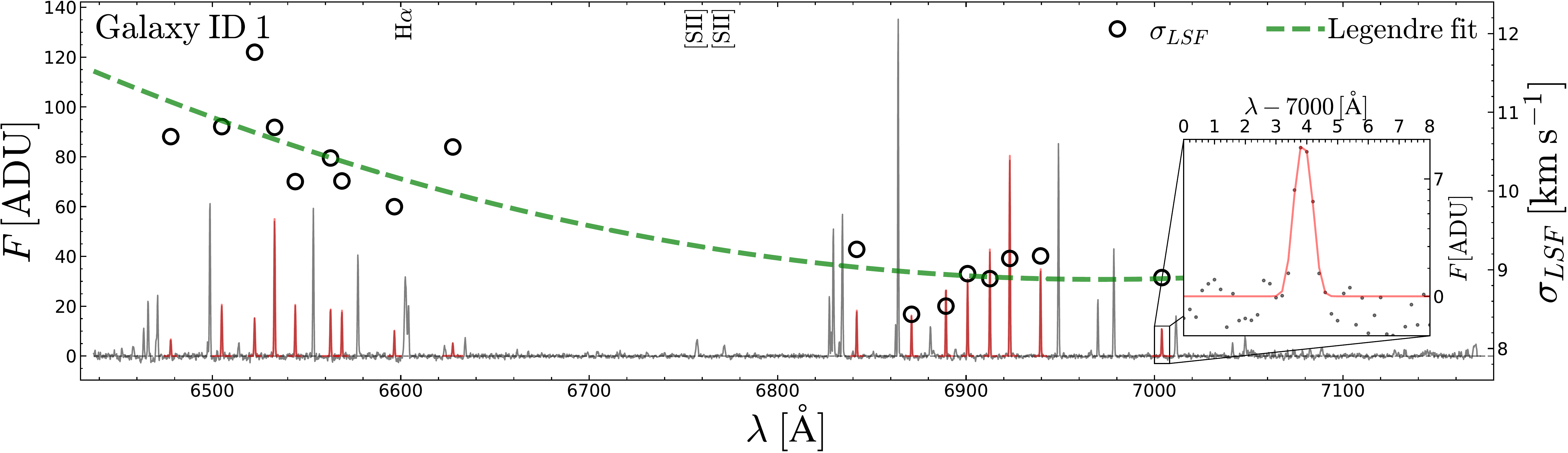}
  \caption{The spectral resolution is well approximated by a second-order polynomial in wavelength. The solid grey line is the IFU-stacked continuum-subtracted spectrum for galaxy ID~1 (left scale). To measure the instrument spectral resolution (right scale), we fit a Gaussian to the most prominent singlet sky lines (best-fit models are overlaid as solid red lines). The measured dispersions are marked by circles and are located at the best-fit wavelength of the relevant sky line. The dashed green line is the best-fit second-order Legendre polynomial to the line dispersions. The interpolated spectral resolution at $\lambda = 6882$\,\AA\ is $\sigma = 9.06$\,km\,s$^{-1}$. The inset diagram shows an example line fit, where the black dots represent the data and the red line traces the best-fit model.}\label{f.sad.specres}
\end{figure*}

\begin{table}
  \setlength{\tabcolsep}{2pt}
  \caption{Prominent singlet sky emission lines used to measure the instrument spectral resolution \citep[from][]{hanuschik2003}.}\label{t.sad.skylines}
  \begin{tabular}[t]{ccc}
  \hline
  Line ID & $\lambda_c \; [\text{\AA}]$ & $\mathrm{FWHM} \; [\text{\AA}]$ \\
  (1) & (2) & (3) \\
  \hline
   1 & 6477.921 & 0.158 \\
   2 & 6505.000 & 0.251 \\
   3 & 6522.433 & 0.166 \\
   4 & 6533.050 & 0.173 \\
   5 & 6544.036 & 0.163 \\
   6 & 6562.760 & 0.226 \\
   7 & 6568.789 & 0.159 \\
   8 & 6596.645 & 0.175 \\
   9 & 6627.632 & 0.252 \\
  \end{tabular}
  \begin{tabular}[t]{ccc}
  \hline
  Line ID & $\lambda_c \; [\text{\AA}]$ & $\mathrm{FWHM} \; [\text{\AA}]$ \\
  (1) & (2) & (3) \\
  \hline
  10 & 6841.963 & 0.154 \\
  11 & 6871.073 & 0.158 \\
  12 & 6889.302 & 0.173 \\
  13 & 6900.808 & 0.157 \\
  14 & 6912.638 & 0.162 \\
  15 & 6923.192 & 0.189 \\
  16 & 6939.542 & 0.150 \\
  17 & 7003.873 & 0.253 \\
  \end{tabular}
\end{table}

The instrument dispersion $\sigma_\mathrm{instr}$ was then calculated from the best-fit line dispersion by subtracting in quadrature the intrinsic dispersion. The results from the skyline measurements are shown in Figure~\ref{f.sad.specres}. The instrument dispersions were approximated across the full wavelength range with a second-order Legendre polynomial,
\begin{equation}
    c_0 + c_1 x + \dfrac{c_2}{2}\left(3 x^2 - 1\right) ~,
\end{equation}
with coefficients $(c_0, c_1, c_2) = (9.58294, -1.1229,  0.4273)$ and with $x = (\lambda - 6740.91)/262.98$ (i.e.\ $x \in (-1,1)$ over the wavelength range considered). With this approximation, the instrument dispersion at $\lambda = 6822$\,\AA\ is $\sigma_\mathrm{inst} = 9.1$\,km\,s$^{-1}$, or 0.487\,\AA\ in terms of FWHM. We adopt this value as the spectral resolution of the \shade\ survey.

As a further test, we measure the arc emission lines after applying the wavelength solution, and find results consistent with the measurements from the sky emission lines. In principle, adding the spectra from different spaxels may artificially broaden the line-spread function, so that the instrument resolution measured from the co-added spectrum might be coarser than the resolution measured from individual spaxels. However, in practice, we find no systematic difference between the line widths measured from the co-added spectra or from individual spaxels (although the latter show larger dispersion, as expected from their lower overall SNR).

\subsubsection{Sky subtraction}

The sky subtraction was performed using the penalised pixel fitting algorithm \citep[\texttt{pPXF};][]{cappellari2017}. As sky templates, we used the spaxels within the ARGUS IFU further away from the target galaxy that belong to the lowest 5\% of the distribution of \halpha flux over the whole IFU footprint.

\begin{figure}
  \centering
  \includegraphics[type=pdf,ext=.pdf,read=.pdf,width=\columnwidth]{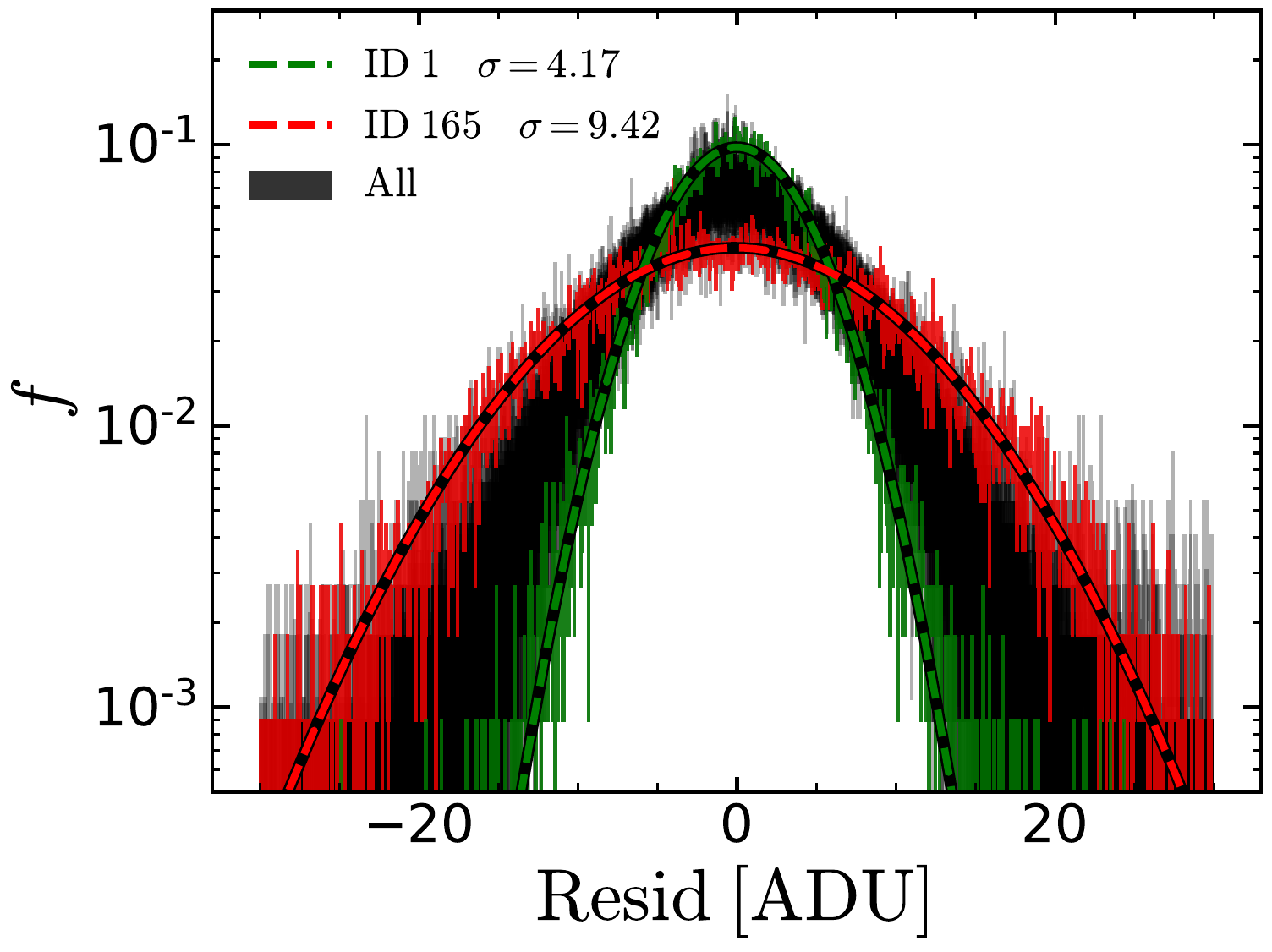}
  \caption{The distribution of sky subtraction residuals for each galaxy. The best and worst cases are highlighted in green and red (galaxy 1 and galaxy 165 respectively). The dashed lines show Gaussian distributions with zero mean and standard deviations from the associated distribution of residuals. The good match between these Gaussians and the histograms suggests that the sky residuals are normally distributed.}\label{f.sad.skyresid}
\end{figure}

The quality of the sky subtraction is assessed by subtracting the best-fit sky from each of the sky spaxels. To avoid a trivial fit, for any given sky spectrum under consideration we remove it from the library of sky templates. The residuals are then computed as the difference between the observed sky spectrum and the best-fit sky. The distribution of the residuals is shown in Figure~\ref{f.sad.skyresid}. This overall distribution is not Gaussian but is composed of residual distributions for individual galaxies that are Gaussian but with different standard deviations. We highlight the best case, with a standard deviation of 4.2\,ADU (galaxy 1; green histogram in Figure~\ref{f.sad.skyresid}), and the worst case, with a standard deviation of 9.4\,ADU (galaxy 165; red histogram in Figure~\ref{f.sad.skyresid}). These two histograms are described very well by Gaussians with zero mean and standard deviation equal to standard deviation of the residuals for that galaxy. For each galaxy, we use test the null hypothesis that the residuals are drawn from a Gaussian using the Kolmogorov-Smirnov test. We adopt a generous significance treshold $p = 0.05$, yet find no galaxy exceeding this limit. For the two examples considered, we find a $p$-value of 0.99; the lowest 
$p$-value is 0.46 for galaxy 496.

\begin{figure*}
  \centering
  \includegraphics[width=1.0\textwidth]{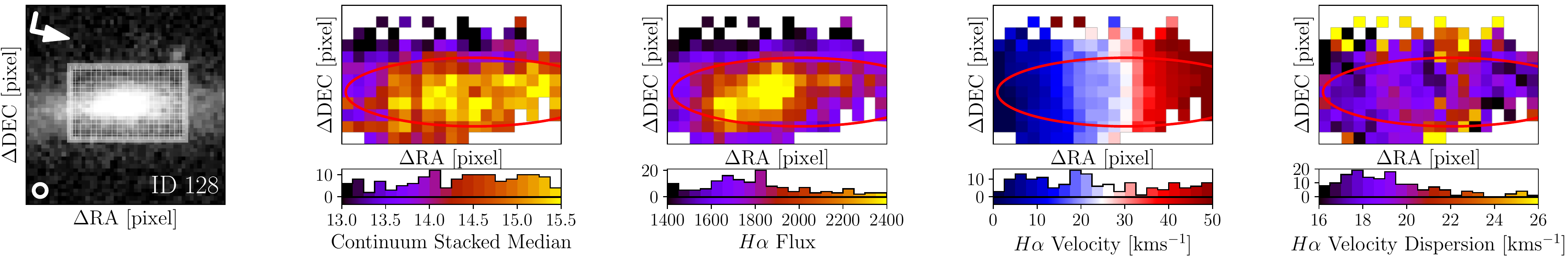}
  \includegraphics[width=1.0\textwidth]{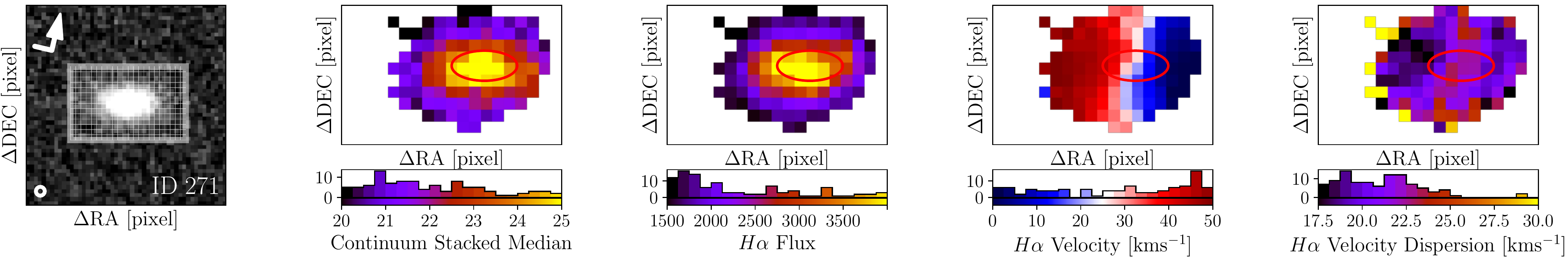}
  \includegraphics[width=1.0\textwidth]{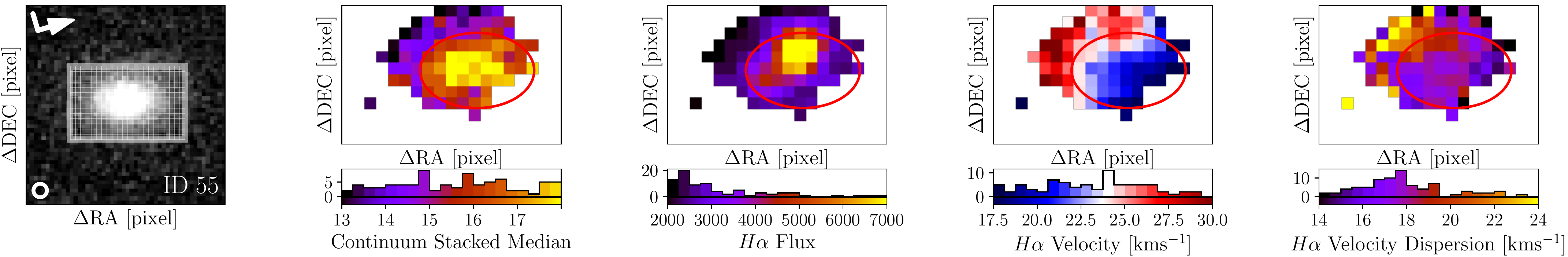}
  \includegraphics[width=1.0\textwidth]{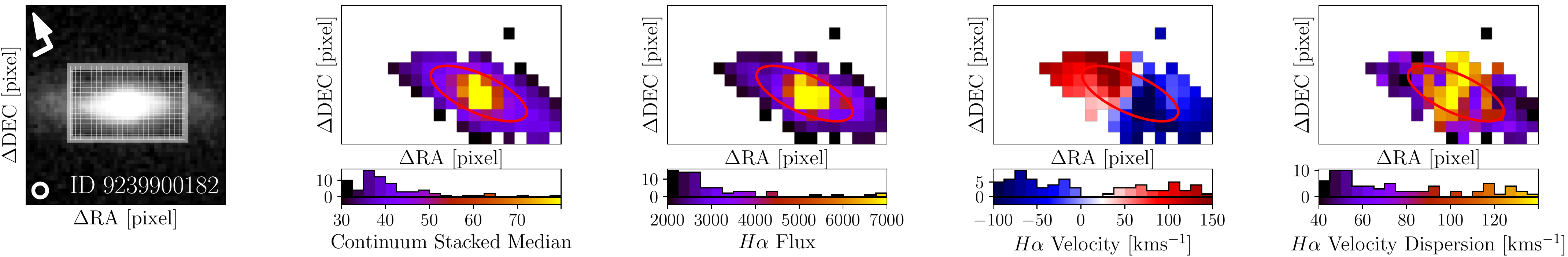}
  \caption{Example data from the \shade\ survey, spanning $0.003 \leq z \leq 0.055$ and $10^{7.43} \leq {M_{\star}} \leq 10^{10.60} \rm M_{\odot}$. From left to right: SDSS $i$-band image; \halpha observed flux map; and \halpha velocity and velocity dispersion maps. SDSS $i$-band images also show the FLAMES-ARGUS instrument target footprint on the galaxy with a grid representing the \shade\ spaxels; the small circles in the $i$-band images have diameter equal to the PSF FWHM in pixel units; each $i$-band image has been rotated to align the galaxy's position angle to the long axis of the ARGUS IFU (the North/East directions are indicated by the white arrow/short arm). Where there is misalignment, we use \texttt{mgefit} to measure the position angle from stacked continuum image for our spaxel sampling. The flux and kinematic maps only show spaxels with SNR > 5; each map also has an associated histogram showing the distribution of the observed quantity. The red ellipse in each \shade\ data map represents the sampling region inside 1$R_e$.}
  \label{fig:kinmap}
\end{figure*}

\subsection{Ancillary data}\label{subsec:ancillarydata}

The ancillary data used in this study includes SDSS DR12 redshift ($z$) and within-fibre star formation rate densities ($\Sigma_\mathrm{SFR}$). The redshifts are emission-line based spectroscopic redshifts. The stellar mass ($M_\star$) was obtained from $i$-band magnitude and $g-i$ colour \citep{taylor+2011}, using the k-correction from \citet{bryant+2014}. $\Sigma_\mathrm{SFR}$ was obtained from the  extinction-corrected \halpha luminosities, divided by the fibre area in physical units. Notice that, because of the \halpha-flux 
selection criterion, the \shade~sample might be biased to higher SFR than the average at its
$M_\star$ (solid green line in Fig.~\ref{f.sad.sample.b}). Higher-than-average SFR cause 
bluer-than-average $g-i$ colour, which could bias our estimated $M_\star$, because this colour features directly in the expression of $M_\star$. To quantify this 
potential bias, we firstly measured the median $g-i$ colour of the \shade~parent sample (i.e.
the sample prior to any morphological or \halpha-flux cut). This colour is $0.48 \, 
\mathrm{mag}$, and is indeed significantly redder than the median colour of the \shade~sample
($0.30 \, \mathrm{mag}$). If we replace the measured $g-i$ colour of the \shade~galaxies with
the median colour of the parent sample, we infer $0.13 \, \mathrm{dex}$ higher $M_\star$.
We remark that this bias is to be considered an upper limit, because it assumes that all the
$i-$band light of the \shade~galaxies is emitted from older stellar populations, as old as the
median age of the parent sample, whereas, in reality, the stellar populations that dominate
the $g-i$ colour also contribute - in part - to the $i-$band light.

For each galaxy, we used $i$-band SDSS DR12 photometry to measure the circularised half-light radius ($R_{e}$), the ellipticity ($\epsilon$) and the position angle using a multi-Gaussian expansion \citetext{MGE, \citealp{emsellem+1994}; we use \texttt{mgefit}\footnote{Available in the \href{https://pypi.org/project/mgefit/}{Python Package Index} (PyPI)}, the python implementation of \citealp{cappellari2002}}. The method is described in D'Eugenio et~al. (in~prep.); here we briefly summarise the main steps. For each galaxy, we retrieve an image from the SDSS database that is 400\,arcsec on a side and centred on the galaxy. We use \texttt{PSFEx} \citep{bertin2011} to identify a set of unsaturated stars from which to measure the point spread function (PSF). The local PSF is then modelled as a sum of two circular Gaussian functions and used as input to \texttt{mgefit}. $R_e$ is calculated from the best-fit MGE model, following the definition of \citet{cappellari2002}; $\epsilon$ is defined as the ellipticity of the model isophote of area $\pi R_e^2$.

Our photometric parameters are in excellent agreement with the measurements from the SAMI Survey, obtained using single-S{\'e}rsic profiles \citep{kelvin+2012, owers+2017}; the root mean square (rms) scatter between the MGE and S{\'e}rsic $R_e$ measurements is 0.06\,dex, implying a rms measurement uncertainty of 0.045\,dex for both GAMA and MGE measurements if distributed equally (D'Eugenio et~al. in~prep.).

\section{Data analysis}
\label{sec:dataanalysis}

This work primarily focusses on the galaxy scaling relations of dwarf galaxies, especially the $\log M_{\star}- \log S_{0.5}$ relation. For this endeavour, the most important parameters are $V_\mathrm{rot}$ and $\sigma$. Since \shade\ is an IFS survey that offers us data cubes with a spectrum at every location within the FOV, we perform single-component emission-line fitting for the \halpha, [N{\sc ii}] and [S{\sc ii}] lines. Then, from the spaxel-level kinematics, we calculate global $V_\mathrm{rot}$ and $\sigma$ values. We also investigate the quality of the spaxel kinematics and look for any biases within them. The following subsections present the information extracted from the analysis in more detail. 

\subsection{Spaxel kinematics}
\label{subsec:spaxkin}

IFS allows us to study the kinematics of gas and stars at each location (spaxel) within a galaxy. For \shade, galaxy gas kinematics are fitted using \texttt{pPXF}, and a set of Gaussian emission-line templates, consisting of the \halpha line and the [N{\sc ii}] and [S{\sc ii}] forbidden lines. Each line is convolved with a Gaussian having standard deviation equal to the instrumental spectral resolution. We use the appropriate spectral resolution values at the wavelength positions of the emission lines from the interpolated Legendre function (see Section~\ref{s.d.ss.dr.sss.wavecal} and the dashed green line in Figure~\ref{f.sad.specres}). In each iteration, \texttt{pPXF} creates a model spectrum by convolving the input templates with a trial velocity dispersion $\sigma$ and applying a trial offset $v$: the best-fit $\sigma_i$ and $v_i$ are those that minimise the $\chi^2$ of the residuals (subscript $i$ here runs over spaxels). 

In this work, we use the \halpha velocity dispersion measurements independently of the [N{\sc ii}] and [S{\sc ii}] forbidden-line measurements. However our \halpha velocity dispersion measurements do not change if we constrain \halpha to have the same kinematics as [N{\sc ii}] and [S{\sc ii}]. Figure \ref{fig:kinmap} shows the SDSS photometry and \halpha flux, velocity, and velocity dispersion maps for four example \shade\ galaxies. From the figure we can see that the \shade\ IFS maps clearly reveal the distributions of \halpha flux, rotation velocity, and velocity dispersion in each galaxy.

 We obtain the uncertainties in $v_i$ and $\sigma_i$ using a Monte Carlo approach. For each spaxel, we create 100 spectra by taking the best-fit spectral model and adding random noise. This noise is the residual between the best-fit model and the observed spectrum around the \halpha line, shuffled in wavelength. We then run \texttt{pPXF} on each of these 100 realisations to estimate the rms uncertainties in the systemic velocities ($\Delta v_i$) and velocity dispersions ($\Delta \sigma_i$).
 
 For each spaxel, we take the SNR of the \halpha line to be 
 \begin{equation}
     SNR=\frac{F}{\sqrt{N}\sigma_{\lambda}} ~,
     \label{eq:SNR}
 \end{equation}
 where $F$ is the integrated flux of the \halpha line, $N$ is the number of pixels the \halpha line spans, and $\sigma_{\lambda}$ is the standard deviation of the residual noise under the \halpha line.

\subsection{Systematic errors}

\begin{figure}
  \centering
  \hspace*{-0.5cm}
  \includegraphics[scale=0.6]{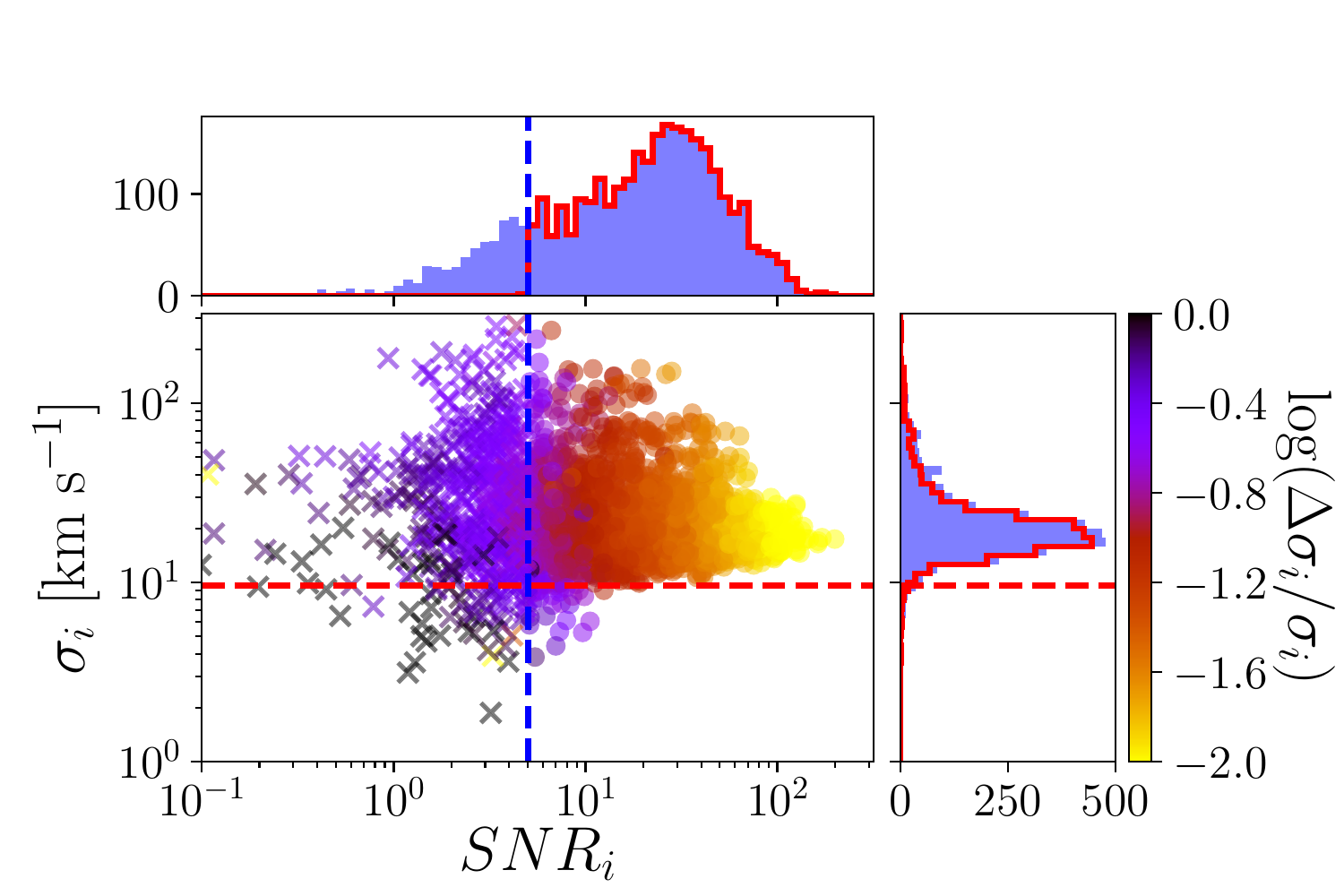}
  \caption{Velocity dispersion quality of all \shade\ galaxy spaxels and the effect of the SNR$_i$\,>\,5 quality cut. The red dashed horizontal line shows the typical spectral resolution of 9.6\,km\,s$^{-1}$ and  the blue dashed vertical line shows the quality cut we applied at SNR$_i$\,>\,5. Circles and crosses are spaxels above and below the SNR cut respectively and spaxels are colour-coded by their relative uncertainties. The blue and red marginal histograms show the distribution of measurements before and after the quality cut, respectively. This plot shows that, with this SNR limit, our spaxel $\sigma$ measurements show no obvious bias and the majority of spaxels have SNR$_i$\,>\,10 and relative uncertainties in $\sigma_i$ below 15\%.}
  \label{fig:spaxelquality}
\end{figure}

Measuring accurate velocity dispersions is notoriously challenging; success depends on a combination of sufficient spectral resolution and SNR. In order to evaluate the possibility of a bias in our measurements of $\sigma_i$, we collect $\sigma_i$, $\Delta \sigma_i$, and SNR$_i$ measurements from all spaxels within an elliptical aperture for each galaxy. The apertures are defined by the half-light radius $R_e$, ellipticity $\epsilon$, and position angle of the galaxies, see red ellipse in Figure~\ref{fig:kinmap} for example. For a few galaxies in our sample, there was misalignment between the IFU positioning and the galaxy position angle, for these cases we used \texttt{mgefit} to measure the position angle from stacked continuum images. Figure~\ref{fig:spaxelquality} shows the distribution of $\sigma_i$ and SNR$_i$ within these apertures for all \shade\ galaxies, colour-coded by the relative error $\log(\Delta\sigma_i/\sigma_i)$. The blue filled histograms in Figure~\ref{fig:spaxelquality} show that the $\sigma_i$ measurements in our sample are approximately symmetric over the range $1.0 < \log\sigma_i < 1.5$ without any significant skew towards over- or under-estimation of velocity dispersion. Moreover, 85\% of spaxels  have SNR\,$\geq$\,5, shown by the vertical dashed line; for our study, we keep all spaxels with SNR$_i\geq5$. A quality cut of SNR$_i\geq5$ results in more than three quarters of spaxels (78\%) having $\log\Delta\sigma_i/\sigma_i \leq -1$. This SNR limit (corresponding to the red open histograms in both marginal distributions) does not introduce any bias in the $\log \sigma_i$ distribution. Note that the distribution of $\sigma_i$ is peaked well above the instrumental spectral resolution, shown by the horizontal dashed line in the figure; the distribution implies we are resolving the internal motions for 92\% of the spaxels in these galaxies. Our main findings do not change with a SNR\,$\geq$\,10 cut. 

\subsection{\texorpdfstring{\boldmath$V_\mathrm{rot}$}{Lg} and \texorpdfstring{\boldmath$\sigma$}{Lg} measurements}
\label{subsec:kin_measure}

To study various kinematic scaling relations, we measure the global $V_\mathrm{rot}$ and $\sigma$ for each galaxy. For these, we follow the approach of \citet{catinella+2005} and \citet{cortese+2014} to remain consistent with \citetalias{barat+2019}. Here we provide a brief overview of the method; for more details see Section~2.2 of \citetalias{barat+2019}. 

For the rotation velocity $V_\mathrm{rot}$, we use the histogram technique where for each galaxy we measure the velocity width ($W$) between the 10th and 90th percentiles of the $v_i$ distribution for spaxels within the 1$R_e$ elliptical aperture and correct for inclination ($i$) and redshift ($z$). Following \citetalias{barat+2019}, we do not perform inclination corrections for nearly edge-on galaxies with (minor-to-major axes ratio) $b/a<0.2$ and we exclude nearly face-on galaxies with $b/a>0.95$.

For the velocity dispersion $\sigma$, we measure the root mean square velocity dispersion from all the spaxels within the 1$R_e$ elliptical aperture, weighted by the continuum flux. 
It is worth noting that both $V_\mathrm{rot}$ and $\sigma$ are calculated using only spaxels with SNR$\geq5$. For the uncertainties in the global quantities, $\Delta V_\mathrm{rot}$ and $\Delta \sigma$ (as well as for $\Delta S_{0.5}$), we use a bootstrap method: we randomly pick the same number of spaxels as the total number within the aperture (allowing repetitions); we calculate $V_\mathrm{rot}$, $\sigma$ and $S_{0.5}$ from these spaxels, as above; and we repeat this 1000 times, using the resulting standard deviations as $\Delta V_\mathrm{rot}$, $\Delta \sigma$ and $\Delta S_{0.5}$. The kinematic parameters for the \shade~sample and the control sample used for this work can be found in Table \ref{tab:long}. 

\begin{figure*}
  \centering
  \includegraphics[width=\textwidth]{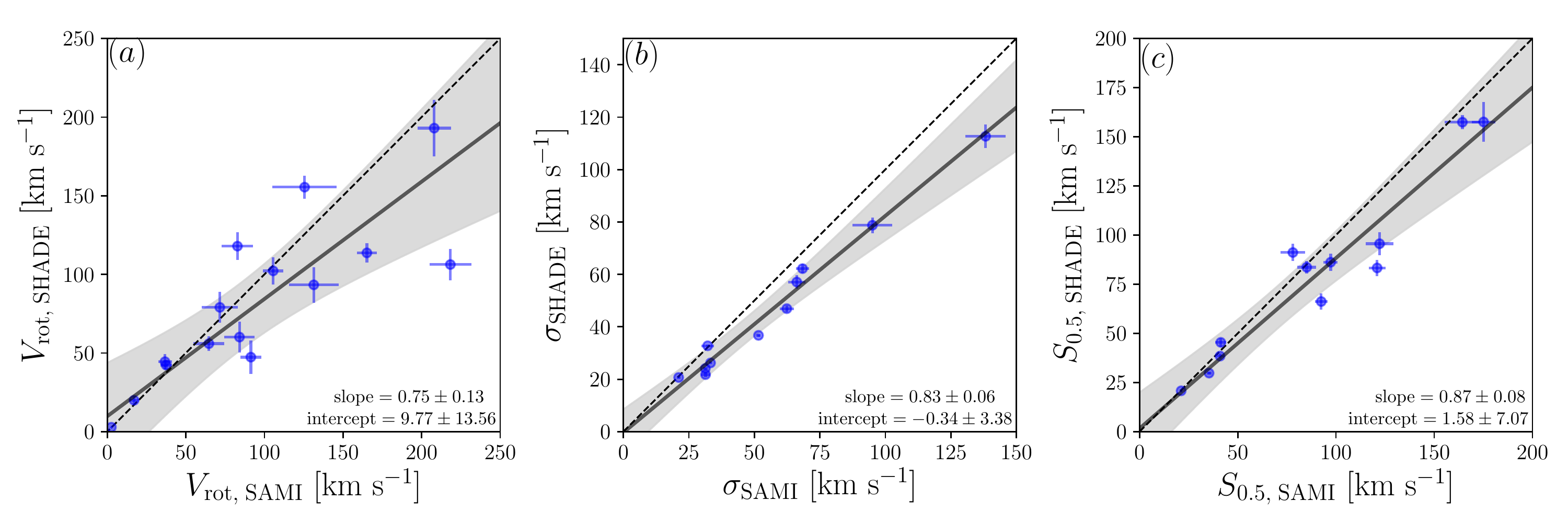}    {\phantomsubcaption\label{fig:S05_comparison.a}
    \phantomsubcaption\label{fig:S05_comparison.b}
    \phantomsubcaption\label{fig:S05_comparison.c}}
  \caption{Comparing SAMI and \shade\ kinematic parameters for the control sample. The dashed lines are one-to-one relations, the solid lines are the linear best fits to the paired quantities, with the shaded region indicating the 99\% confidence interval in the quantity on the vertical axis given the uncertainties in the fitted slope and intercept. The measured $\sigma$ values are lower for \shade\ than SAMI while $V_\mathrm{rot}$ and $S_{0.5}$ are consistent between \shade\ and SAMI.}
  \label{fig:S05_comparison}
\end{figure*}

\begin{figure*}
  \centering
  \includegraphics[width=\textwidth]{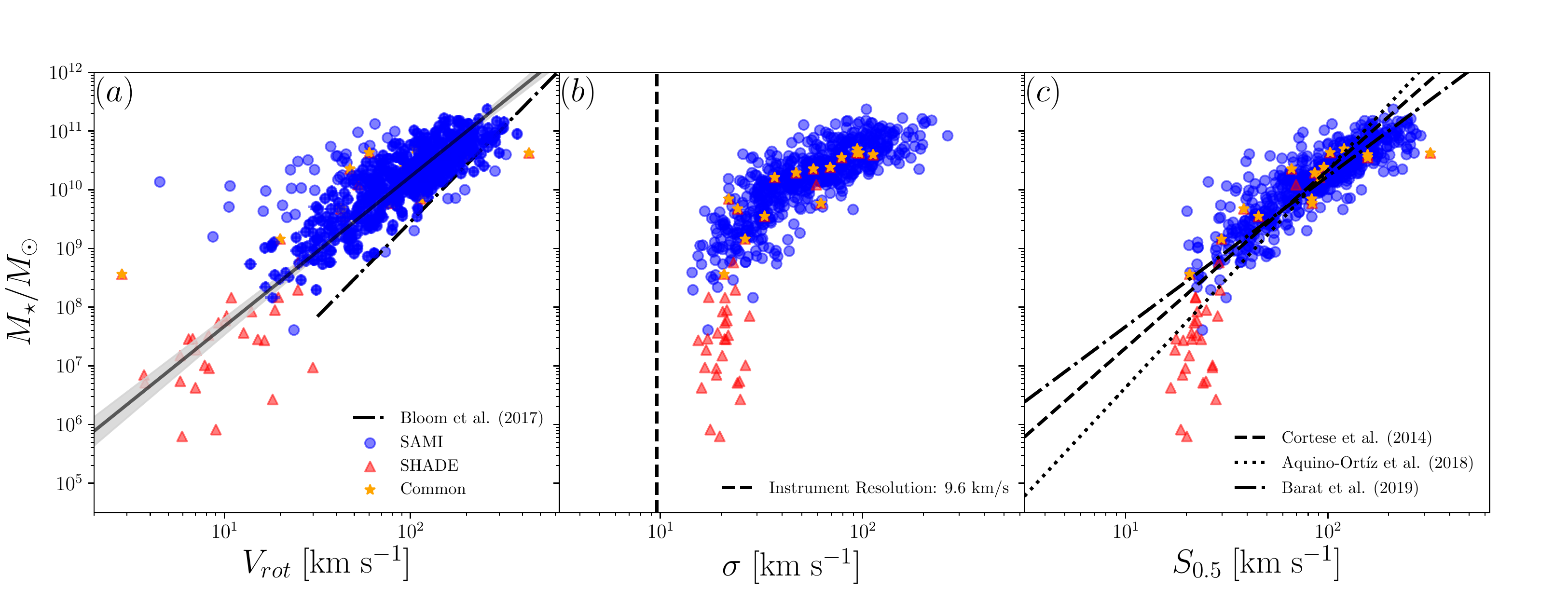}
  {\phantomsubcaption\label{fig:TFFJS05_comparison.a}
    \phantomsubcaption\label{fig:TFFJS05_comparison.b}
    \phantomsubcaption\label{fig:TFFJS05_comparison.c}}
  \caption{SAMI and \shade\ Tully-Fisher (TF), Faber-Jackson (FJ), and $S_{0.5}$ scaling relations. Blue circles represent SAMI measurements, red triangles represent \shade\ measurements, and yellow stars represent \shade\ measurements of the control sample from SAMI. In the TF relation shown in panel~(a), the \shade\ sample in general follows the same relation (solid line) as the SAMI sample. The shaded region in panel~(a) represents the 99\% confidence interval in the vertical axis given the uncertainties in the fitted slope and intercept. The combined TF relation is flatter than that obtained by \citet{bloom+2017b}. The FJ relation in panel~(b) shows a clear bend and a lower limit for the $\sigma$ measurements above the spectral resolution limit of the \shade\ observations (shown by the vertical dashed line). The bend in the FJ relation carries across to the $S_{0.5}$ scaling relation in panel~(c). Best-fit lines from \citet{barat+2019,Aquino-Ortiz2018,cortese+2014} are included for comparison in panel~(c).}
    \label{fig:TFFJS05_comparison}
\end{figure*}

\section{Scaling relations}
\label{s.res}

After obtaining the $S_{0.5}$ kinematic parameter, we begin the analyses of kinematic scaling relations. As the \shade\ sample contains 20 galaxies that are in common with the SAMI survey, we first perform a comparison between SAMI and \shade\ kinematic measurements to determine if there is any systematic bias in any of samples. Then we construct the TF, FJ and the $\log M_{\star}$--$ \log S_{0.5}$ scaling relations to investigate how dwarf galaxies behave on them. In this section we also compare the stellar and baryonic versions of the $\log M - \log S_{0.5}$ scaling relation.

\subsection{Comparing \texorpdfstring{\boldmath\shade}{Lg} and SAMI kinematics}
\label{subsec:SHADE&SAMIKIN}

After obtaining the gas kinematics of the \shade\ galaxies, we compare our measurements with those from the SAMI survey. Of the 69 galaxies observed in SHADE, 20 overlap with the SAMI galaxy sample. These galaxies in common are chosen from the SAMI $\log M_{\star}$--$ \log S_{0.5}$ scaling relation \citepalias{barat+2019} to be the control sample (see Section~\ref{s.d.ss.sample}). The control sample has a mass range from $10^{8.6}M_{\odot}$ to $10^{10.6}M_{\odot}$. Because accuracy in velocity dispersion measurements is crucial in our study, these independent measurements of velocity dispersion for the same galaxies provide a critical test of the systematics associated with each survey.

We compare SAMI and \shade\ measurements of $V_\mathrm{rot}$, $\sigma$, and $S_{0.5}$ parameters in Figure~\ref{fig:S05_comparison}. For each plot, we kept only galaxies with $\Delta V_\mathrm{rot}/V_\mathrm{rot}<0.2$, $\Delta \sigma/\sigma<0.1$, and $\Delta S_{0.5}/S_{0.5}<0.1$ respectively. Separate relative error thresholds are chosen to ensure an adequate number of galaxies over sufficient range of values remain in the comparison, while rejecting outliers. The fit parameters in each plot indicate that \shade\ velocity dispersions are consistently lower than SAMI velocity dispersion. This can be explained by a combination of two improvements: firstly, the spectral resolution of \shade\ (9.6~\kms; Section~\ref{s.d.ss.dr.sss.wavecal}) is three times better than SAMI \citep[29.9~\kms;][]{vandesande+2017}, and secondly the spatial resolution of \shade\ is also better, which combined with the improved seeing condition for the \shade  observations, mitigates the effect of beam smearing (see Section~\ref{s.d.ss.obs}). The difference between the \shade and SAMI velocity dispersions is not highly significant ($\sim 2.8\sigma$) and, when combined with the rotation velocity, the $S_{0.5}$ measurements have an insignificant ($\sim 1.6\sigma$) difference. This concordance demonstrates the robustness of the $S_{0.5}$ kinematic parameter against atmospheric seeing. 

\subsection{Kinematics scaling relations of dwarf galaxies}

We extend the kinematic scaling relation studies of \cite{cortese+2014}, \cite{Aquino-Ortiz2018} and \citetalias{barat+2019} to dwarf galaxies by combining \shade\ data and SAMI data from \citetalias{barat+2019} to construct the stellar mass TF, FJ and $S_{0.5}$ kinematic scaling relations over the mass range $10^{5.7} < M_\star < 10^{11.4} \rm M_{\odot}$. As we have shown in Section~\ref{subsec:SHADE&SAMIKIN}, the \shade\ $V_\mathrm{rot}$, $\sigma$, and $S_{0.5}$ measurements are in good correlation with those measured from SAMI data for the control (higher-mass) sample, with small scatter and slight offset. Given this agreement, it is not surprising that the scaling relations from \shade\ connect well with SAMI scaling relations without any obvious offset, as shown in  Figure~\ref{fig:TFFJS05_comparison}. The \shade\ measurements for the control sample (star symbols in Figure~\ref{fig:TFFJS05_comparison}) lie within the SAMI sample distribution in each scaling relation, so there is no need to calibrate the \shade\ and SAMI scaling relations (and indeed our results do not change if we calibrate the kinematic measurements using the results from Section~\ref{subsec:SHADE&SAMIKIN}).

The extended TF relation and the best fit (solid) line to the SAMI sample in Figure~\ref{fig:TFFJS05_comparison.a} show that the dwarf galaxies in \shade\ follow the SAMI TF relation, albeit with greater scatter; fitting both samples simultaneously produces the same line within the uncertainties. For comparison, we included the best fit (dot-dashed) line from \cite{bloom+2017b}, which also uses SAMI data. Our TF relation only agrees with \cite{bloom+2017b} at high masses ($M_\star>10^{10}M_{\odot}$). This difference is due to different sampling regions of the galaxies in the two studies: in \cite{bloom+2017b}, $V_\mathrm{rot}$ is measured from regions out to $2.2R_e$, whereas we sample within $1R_e$. Therefore our results only agree for high-mass galaxies with steep rotation curves, where maximum rotation velocities can occur within $1R_e$ \citep{Yegorova+Salucci2007}.

One of the main motivations of this study is to observe galaxies with $\sigma$ below the spectral resolution of SAMI ($\sim$30~\kms) with higher resolution and so to constrain the FJ and $S_{0.5}$ scaling relations for low-mass galaxies. Our results in Figure~\ref{fig:TFFJS05_comparison.b} show that, despite having an instrumental resolution of 9.6~\kms, low-mass dwarf galaxies in \shade\ do not reach velocity dispersion below $\sim$15~\kms; i.e.\ the distribution of velocity dispersions in these galaxies has a physical (not instrumental) lower limit. For low-mass \shade\ galaxies with $M_\star \lesssim 10^8 \rm M_{\odot}$ (i.e.\ excluding the control sample), the mean velocity dispersion in galaxies is 22$\pm$5~\kms. 

The lower limit for $\sigma$ propagates to the $S_{0.5}$ scaling relation in Figure~\ref{fig:TFFJS05_comparison.c}. All \shade\ galaxies (apart from the control sample), lie beneath the best-fit line for the SAMI sample from \citetalias{barat+2019}. This confirms the bend in the gas $S_{0.5}$ scaling relation observed in \citetalias{barat+2019} and suggests that a linear relation is not adequate to describe the  $\log M_{\star}$--$ \log S_{0.5}$ scaling relation. The exact location of the bend and its implications are discussed in the following section.

\begin{table*}
\centering
\caption{Scaling relation fitting results from this work and \citetalias{barat+2019}.}
 \begin{tabular}{lcccccc}
  \hline\hline
     Source                                     & $\rm Y$  & Slope  & Intercept  &  $ S_{0.5, \rm lim}$\kms & $ Y_{\rm lim}/M_{\odot}$ & Scatter ($\rm MAD_{orth}$) \\ \hline
     Figure \ref{fig:S05plot}                  & $M_{\star}$ & $2.58 \pm 0.02$ & $5.16 \pm 0.05$ & $22.4\pm 1.1$ & $10^{8.64}$ & 0.063 $\pm$ 0.003 \\
     Figure \ref{fig:baryonicS05plot}          & $M_{b}$      & $2.48 \pm 0.02$ & $5.66 \pm 0.09$ & $23.3\pm 1.0$ & $10^{9.05}$ & 0.096 $\pm$ 0.004 \\  
     Figure \ref{fig:HaloMass_ScalingRelation} & $M_{h}$  & $2.22 \pm 0.07$ & $7.84 \pm 0.75$ & - & - & -\\
    \citetalias{barat+2019}, inverted       & $M_{\star}$  & $2.56 \pm 0.01$ & $5.10 \pm 0.05$ & $22.4\pm 1.0$ & $10^{8.56}$ & 0.063 $\pm$ 0.003 \\\hline

 \end{tabular}
 
 \label{table:fittingresult}
\end{table*}

\subsection{A closer look at the \texorpdfstring{\boldmath$S_{0.5}$}{Lg} scaling relation}

\begin{figure}
    \hspace*{-0.2cm}
    \includegraphics[scale=0.5]{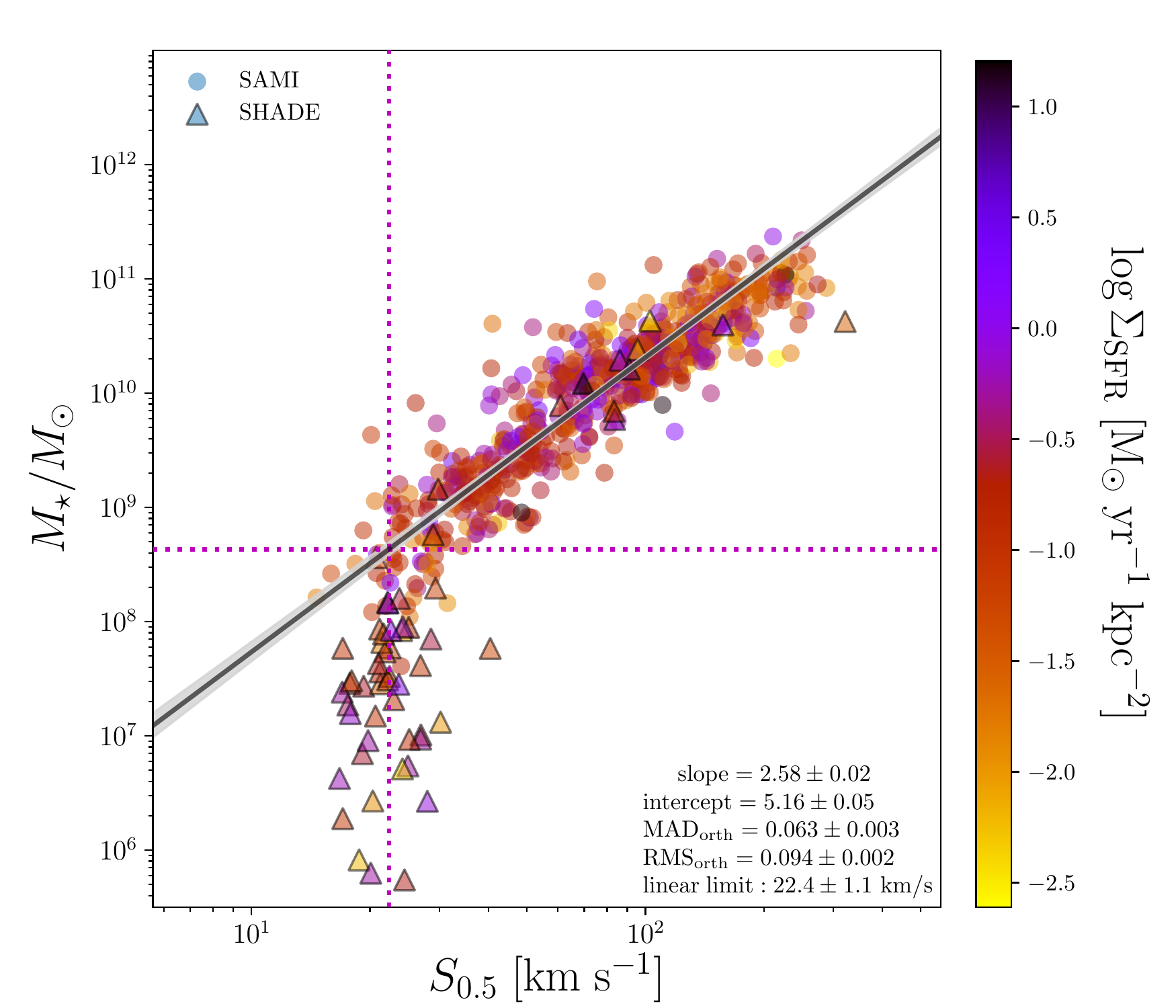}
    \caption{
  $\log M_{\star}$--$ \log S_{0.5}$ scaling relation from combining the SAMI and \shade\ samples. Round points are SAMI measurements; triangular points are \shade\ measurements; all points are colour-coded by gas surface density. The black solid line is the best fit and the shaded region represents the 99\% confidence interval in the vertical axis given the uncertainties in the fitted slope and intercept. Magenta horizontal and vertical lines indicate the limit of the linear fit; below this mass threshold, points are modelled as normally distributed around the vertical line.}
  \label{fig:S05plot}
\end{figure}

\begin{figure}
    \includegraphics[scale=0.6]{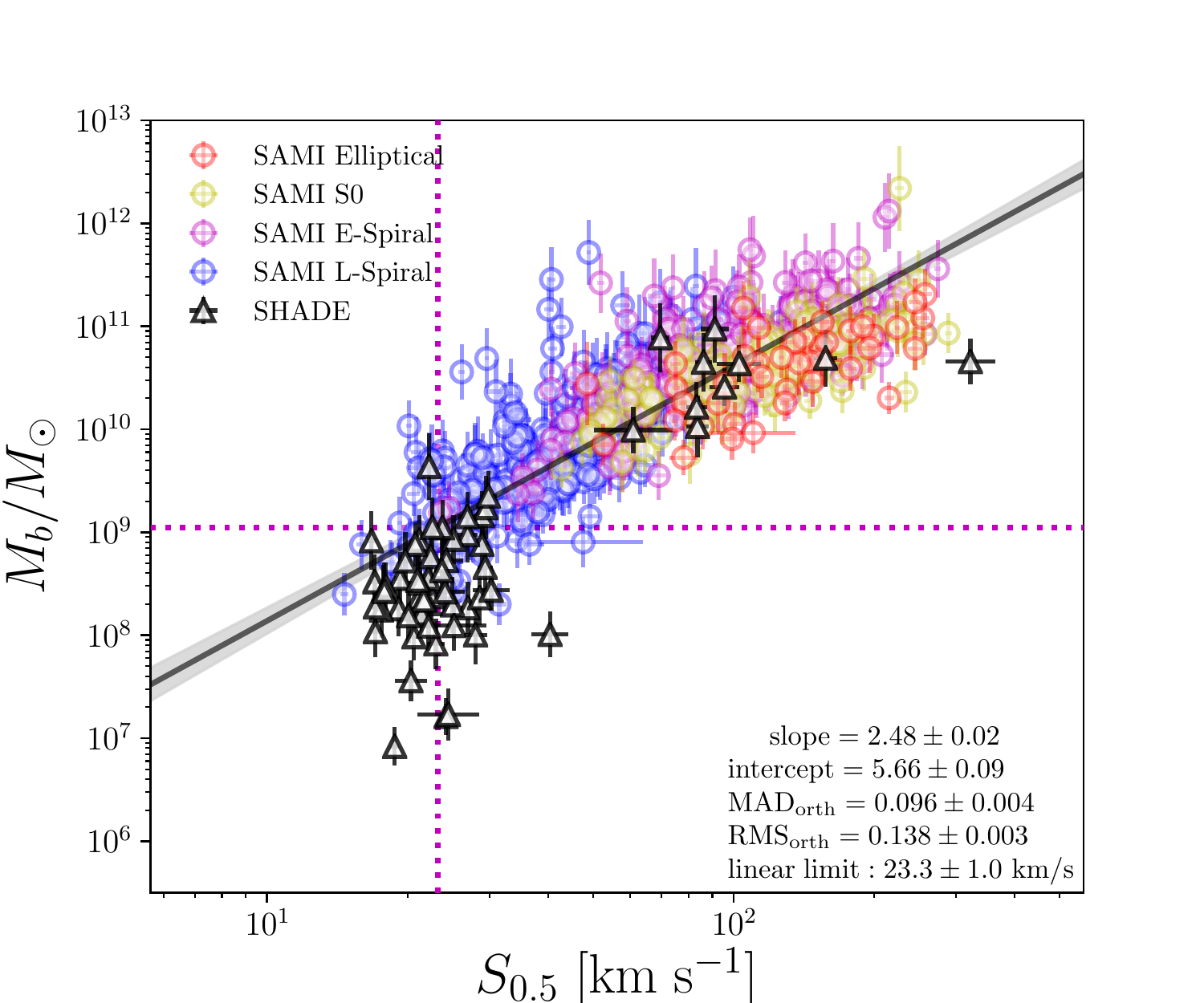}  
    \caption{
    The baryonic version of the $\log M_{\star}$--$ \log S_{0.5}$ scaling relation from combining the SAMI and \shade\ samples. Round points are SAMI measurements; triangles are \shade\ measurements. The molecular and atomic gas mass is estimated based on \citet{Kennicutt1998}. SAMI galaxies are colour-coded by their visual morphologies as elliptical, S0, early-spiral or late-spiral. It is apparent the approximated baryonic scaling relation significantly reduce the bend in the scaling relation.}
  \label{fig:baryonicS05plot}
\end{figure}

Using the combined SAMI and \shade\ galaxy sample, we construct the $\log M_{\star}$--$ \log S_{0.5}$ scaling relation in Figure~\ref{fig:S05plot}. In order to assess the presence of a bend in the scaling relation we adopted a Bayesian approach. We model the data as a mixture of a line with Gaussian scatter above some mass threshold $\log M_{\star,{\rm lim}}$ (corresponding to $\log S_{0.5,{\rm lim}}$ on the scaling relation) and below this mass threshold the $\log S_{0.5}$ values are distributed as a Gaussian about $\log S_{0.5,{\rm lim}}$ (i.e.\ $\log M_\star$ is not determined by $\log S_{0.5}$ for masses below $\log M_{\star,{\rm lim}}$). We adopt flat, uninformative priors on all the fitting parameters: the slope and intercept of the linear relation, the Gaussian scatter about the linear relation ($\log M_{\star} > \log M_{\star,{\rm lim}}$), and the Gaussian scatter about $\log S_{0.5,{\rm lim}}$ ($\log M_{\star} < \log M_{\star,{\rm lim}}$) and the mass threshold itself. We estimate the posterior distribution using a Markov Chain Monte Carlo approach \citetext{\citealp{metropolis+1953}; for details see \citetalias{barat+2019},  Section~3.2.1}. We take the model with the maximum log likelihood to be the best-fit model. The model fitting results are shown in Table \ref{table:fittingresult}. The slope and the intercept are not affected by the addition of dwarf galaxies, because they are the same values we obtained in \citetalias{barat+2019} using only SAMI galaxies (their Table~2, sample~B1, c.f. slope of $2.56\pm0.01$, intercept $5.10\pm0.05$ after inversion for consistency). The linear limit fitted to the combined SAMI+\shade\ data is $S_{0.5,{\rm lim}}=22.4\pm1.1$~\kms, corresponding to a stellar mass limit $M_{\star,\rm{lim}}=10^{8.6}\rm M_{\odot}$, and is consistent with that of \citetalias{barat+2019} (c.f. $S_{0.5,{\rm lim}}=22.4\pm1.0$~\kms). This is interesting because, although a bend in the scaling relation was observed in \citetalias{barat+2019}, we could not rule out the possibility that it was an observational artefact, as the limit (22~\kms) was slightly less than the spectral resolution limit of the SAMI instrument (30~\kms). However this is definitely not the case for \shade\ data, which was observed with a spectral resolution of 9.6~\kms; the fitted limit at 22~\kms is more than twice the \shade\ instrumental resolution and so would appear to be physical. On the other hand, the bend observed by \citetalias{barat+2019} at about 60~\kms in the {\em stellar} scaling relation (as opposed to this {\em gas} scaling relation) may still be an artefact due to the SAMI instrumental resolution limit, which for stars was about 70~\kms. It is very unlikely that the observed bend is due to bias in the determination of $M_\star$ because, along the 
 direction of $M_\star$, \shade galaxies are offset from the linear relation by $1 \,
\mathrm{dex}$ or more; in contrast, we estimate the bias in $M_\star$ to be $\lesssim 0.13 \, 
\mathrm{dex}$ (Section \ref{subsec:ancillarydata}).

\subsection{Baryonic \texorpdfstring{\boldmath$\log M_{b}$--$ \log S_{0.5}$}{Lg} scaling relation} 
\label{subsec: baryonicscalingrelation}

The $\log M_{\star}$--$ \log S_{0.5}$ scaling relation follows from the virial theorem if $M_\star$ is a fixed fraction of the total mass $M_{\rm {total}}$. However, if the $M_\star/M_{\rm {total}}$ ratio varies due to an increased gas fraction in lower mass galaxies \citep[as expected over our wide mass range; see][]{Foucaud+2010}, this will introduce a curvature in the $\log M_{\star}$--$ \log S_{0.5}$ relation. To improve the coupling with $M_{\rm {total}}$, therefore, it is in principle better to include the gas mass by using baryonic mass $M_b$ rather than stellar mass $M_\star$. However, due to the lack of direct \HI\ observations for \shade\ galaxies, we have to construct the $\log M_{b}$--$ \log S_{0.5}$ scaling relation by summing the stellar mass with an approximate estimate of the gas mass, which we derive for each SAMI and \shade\ galaxy from its star formation rate (SFR) as follows. The SFR is obtained from SDSS where available (see Section~\ref{subsec:ancillarydata}) and converted to SFR surface density ($\Sigma_{\rm SFR}$) by dividing by the SDSS fibre aperture in kpc$^2$ at the redshift of the galaxy. From $\Sigma_{\rm SFR}$, we then estimate the surface density of neutral and molecular hydrogen gas ($\Sigma_{\rm gas}$) by inverting the star formation law \citep{Kennicutt1998}, described by:
\begin{equation}
    \Sigma_{\rm{SFR}} = \left( 2.5\pm0.7 \right)\times10^{-4} 
    \left(\frac{\Sigma_{\rm{gas}}}{\rm{M_{\odot}\,pc^{-2}}}\right)^{1.4\pm0.15}
    \rm{M_{\odot}\,yr^{-1}\,kpc^{-2}} ~.
    \label{eq:KS98}
\end{equation}
We also explored other, more recent, variants of the star formation law, namely those of \citet{Federrath2017} and \citet{delosReyes+Kennicutt2019}. However the \citet{Federrath2017} relation only estimates cold molecular gas while the \citet{delosReyes+Kennicutt2019} uses UV-based estimation, requiring a conversion and thus introducing additional uncertainties and systematics. We therefore choose to employ the original star formation law from \citet{Kennicutt1998}, which relates \HI~and CO densities to \halpha SFRs.

Once we have obtained $\Sigma_{\rm gas}$, we multiply this gas surface density by the projected area of the galaxy defined by $R_e$ (in kpc$^2$) and ellipticity to obtain the gas mass $M_{\rm gas}$, which we sum with $M_\star$ to get the estimated total baryonic mass $M_b$. For the uncertainties in our $M_{b}$ measurements, we use the uncertainties in Equation~\ref{eq:KS98} and perform Monte Carlo sampling 1000 times. This provides a distribution of $M_b$ for each galaxy and we take the standard deviations as uncertainties.

Figure~\ref{fig:baryonicS05plot} shows the baryonic version of the $S_{0.5}$ scaling relation, with $M_b$ obtained using Equation~\ref{eq:KS98}. The fitting results are shown in Table \ref{table:fittingresult}. Comparing Figure~\ref{fig:S05plot} and Figure~\ref{fig:baryonicS05plot}, we can see that the inclusion of an $M_{\rm gas}$ estimate reduces the extent of the bend in the relation at low masses: the ratio between $M_{\star,\rm{lim}}$ and the observed stellar mass is more than two orders of magnitude in Figure~\ref{fig:S05plot}, whereas, for $M_b$, the corresponding ratio is approximately one order of magnitude (Figure~\ref{fig:baryonicS05plot}). This improvement is achieved by compressing the baryonic mass for dwarf galaxies at the expense of increased scatter at all masses. There is significant overlap between \shade\ and SAMI galaxies in the $\log M_{b}$--$ \log S_{0.5}$ relation over the range $10^8$--$10^9\,\rm{M_\odot}$, with very few galaxies remaining below $10^7\,\rm{M_\odot}$. While the intercept and the scatter of the scaling relation increased as expected, since more mass (and uncertainty) is added, the slope of the relation remains approximately the same (within one standard deviation). Although visually the bend in $\log M_{b}$--$ \log S_{0.5}$ has been reduced by the addition of the gas mass, the linear-regime limit from the model fits indicates that the bend is still present. This suggests that using baryonic mass in dwarf galaxies is still not sufficient to account for their gas kinematics, in contrast to their higher mass ($M_b > 10^{10}\,\rm{M_\odot}$) counterparts. Moreover, we note that, compared to the stellar mass scaling relation, the baryonic mass scaling relation appears to flatten out at masses $M_b > 10^{11}\,\rm{M_\odot}$, which could be due to the inapplicability of the star formation law at the high-mass end the of the scaling relation, where the hydrogen gas converts from atomic to molecular form \citep{bigiel+2008}.

\begin{figure*}
    \centering
    {\phantomsubcaption\label{fig:SFR_Mbar_plot.a}
    \phantomsubcaption\label{fig:SFR_Mbar_plot.b}}
    \includegraphics[scale=0.85]{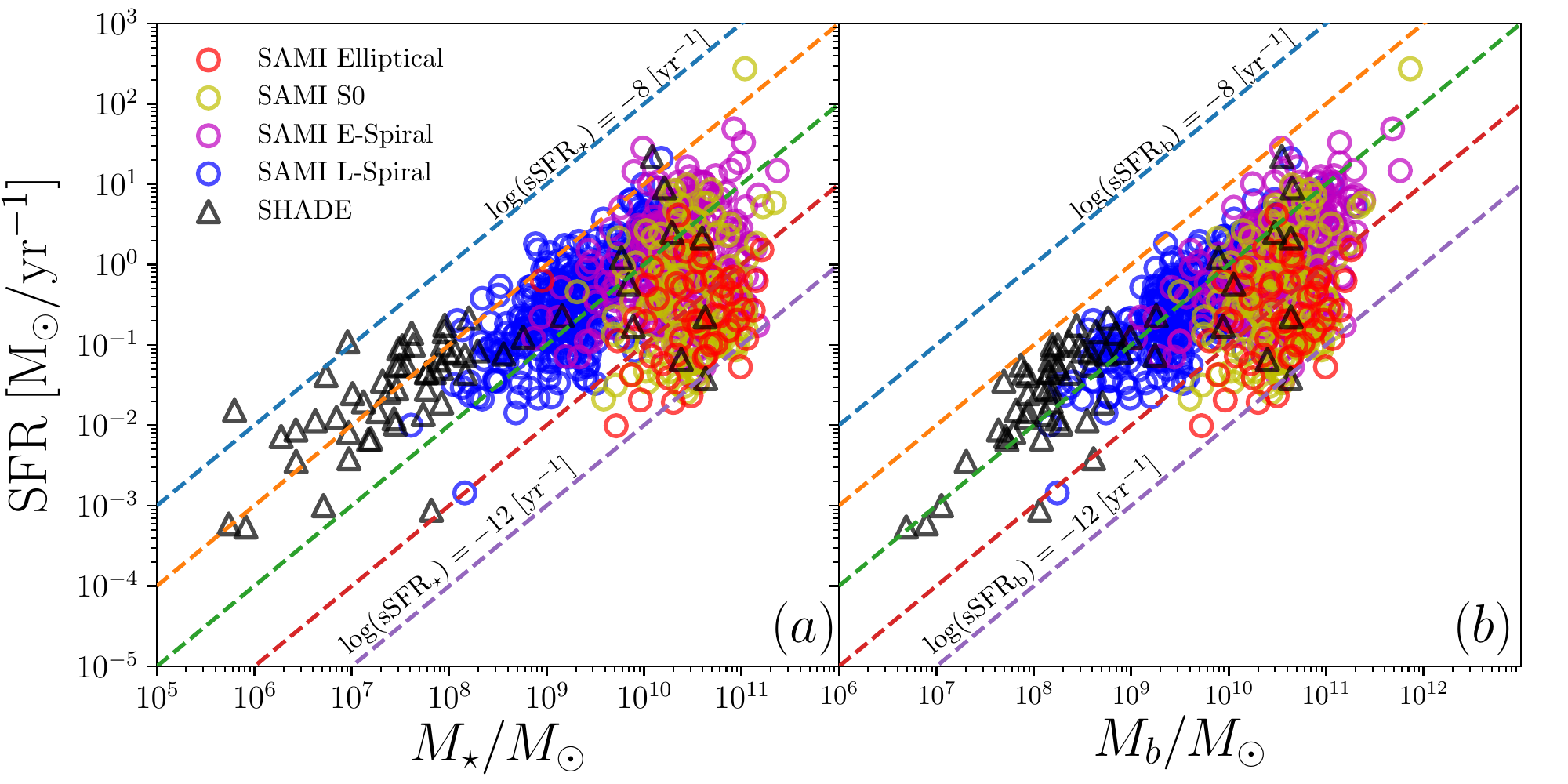}
    \caption{The star-forming sequence of the combined SAMI and \shade\ sample. Panel~(a) shows SFR as a function of stellar mass; the dashed lines indicate specific SFR (sSFR$_{\star}$) in the range $[10^{-8}, 10^{-12}]\,\rm yr^{-1}$, where sSFR$_{\star}$ is SFR divided by stellar mass. Panel~(b) shows the SFR as a function of baryonic mass; the dashed lines indicate sSFR$_b$ (SFR divided by baryonic mass). The marker shapes and colour scheme are the same as in Figure~\ref{fig:baryonicS05plot}. Comparing panels~(a) and~(b) shows that dwarf galaxies have higher sSFR$_{\star}$ than more massive galaxies, but that this difference largely disappears for sSFR$_b$.}
  \label{fig:SFR_Mbar_plot}
\end{figure*}

\section{Discussion}\label{s.dis}

In this section we discuss our findings regarding the $\log M_{\star}$--$ \log S_{0.5}$ scaling relation, especially the observed lower limit on $\sigma$. We look at the range over which the $\log M_{\star}$--$ \log S_{0.5}$ scaling relation remains linear and the location where the relation bends and becomes steeper. We explore the halo mass version of the scaling relation by taking some simple assumptions. Finally we compare our observed lower limit on $\sigma$ to those in the literature and note several possible drivers for the observed limit.

\subsection{Limitations of the \texorpdfstring{\boldmath$S_{0.5}$}{Lg} scaling relation}

The purpose of constructing the $\log M_{\star}$--$ \log S_{0.5}$ relation is to combine $V_{\rm rot}$ and $\sigma$ into a single kinematic parameter that allows star-forming and quiescent galaxies to be put on a common mass--kinematics scaling relation that is tighter than the stellar mass TF relation and has slope and intercept close to the FJ relation. \citet{cortese+2014} demonstrated that galaxies of all morphologies can be brought onto the same $\log M_{\star}$--$ \log S_{0.5}$ scaling relation; this was confirmed by \citet{Aquino-Ortiz2018} and \citet{Gilhuly+2019} using the CALIFA survey and by \citetalias{barat+2019} using a larger SAMI data set. While these findings pointed towards $S_{0.5}$ possibly providing a universal mass proxy, \citetalias{barat+2019} also showed that there existed an apparent bend in both the gas and the stellar versions of the scaling relation at low masses. However, due to limitations in the S/N ratio, the instrumental resolution, and the sample selection, as well as the fact that the bend occurred at different stellar masses for the gas and stellar scaling relations, the apparent bends found by \citetalias{barat+2019} were arguably observational artefacts.

The \shade\ survey is partly motivated by the question of whether there is a physical component to the low-mass bend in the $\log M_{\star}$--$ \log S_{0.5}$ scaling relation. By observing the \halpha kinematics of low-mass dwarf galaxies using a high spectral resolution instrument, the results from this work show that there does indeed exist a physical limit where the {\em gas} version of the $\log M_{\star}$--$ \log S_{0.5}$ scaling relation no longer follows a linear trend.

We have also seen that the linearity limit in the scaling relation is very close to the floor in $\sigma$ measurements at $\sim$20\,\kms, while $V_{\rm rot}$ within 1$R_e$ had very little contribution. Looking at the TF relation obtained in Figure \ref{fig:TFFJS05_comparison.a}, dwarf galaxies do not significantly deviate from the the TF relation formed by more massive galaxies. This suggests that, depending on the difference in the steepness of the rotation curves and the difference in maximum rotation velocities between massive galaxies and dwarf galaxies, measuring rotation velocities at radii beyond one effective radius could possibly increase the contribution of $V_{\rm rot}$ in the $S_{0.5}$ parameter. In fact, in Figure~\ref{fig:TFFJS05_comparison.a} we can see that the TF relation from \citet{bloom+2017b} measured at 2.2$R_e$ shows an offset from the TF relation measured at 1$R_e$. By observing over a larger area, it is likely the limit of $S_{0.5}$ will shift away from the $\sigma$ floor and potentially decrease the bend in the scaling relation. The role of aperture size will be investigated in future work.

To assess whether dwarf galaxies can be brought onto the linear $\log M_{\star}$--$ \log S_{0.5}$ scaling relation of more massive galaxies, we constructed the baryonic version of the scaling relation. Accounting for the gas mass derived from the star formation rate (SFR) significantly increased the scatter of the relation, but also substantially reduced the severity of the low-mass bend in the scaling relation (though it did not eliminate it). We therefore take a closer look at the star-forming sequence of SAMI and \shade\ galaxies. In Figure \ref{fig:SFR_Mbar_plot}, we plot galaxy SFR against $M_\star$ and $M_b$, and overlay contours of specific SFR (sSFR). Note that in Figure \ref{fig:SFR_Mbar_plot.a}, sSFR$_\star$ is defined as the ratio of SFR to $M_\star$; in Figure \ref{fig:SFR_Mbar_plot.b}, sSFR$_b$ is defined as the ratio of SFR to $M_b$. We can see in Figure \ref{fig:SFR_Mbar_plot.a} that dwarf galaxies reside at or above $\rm sSFR_\star = 10^{-9}\,yr^{-1}$, while more massive star-forming galaxies in SAMI have sSFR$_\star$ below this value. This offset in sSFR$_\star$ disappears if instead we use sSFR$_b$. In Figure \ref{fig:SFR_Mbar_plot.b}, all dwarf galaxies have migrated below the $\rm sSFR_b = 10^{-9}\,yr^{-1}$ line and there are no galaxies of any mass observed above this value of sSFR$_b$. Note that we refrain from calling $\rm sSFR_{b} = 10^{-9}\,yr^{-1}$ an upper limit to the specific star formation rate, because the comparison contains (by construction) an internal correlation between the SFR and $M_b$; we only use this comparison to showcase the consistency between dwarf galaxies and high-mass star-forming galaxies. This suggests that when low-mass galaxies ($M_{\star}<10^9\,{\rm M_\odot}$) are to be included in the scaling relation, $M_b$ is a better proxy than $M_\star$ for dynamical mass across the transition in dynamical regime from star-dominated to gas-dominated galaxies. 

\subsubsection{Caveats regarding gas mass estimations} 

While ideally direct measurements of the \HI~gas mass are required for an observational construction of the baryonic $\log M_{b}$--$ \log S_{0.5}$ scaling relation, in the absence such observations, we have used gas masses estimated using indirect relations given in the literature.
A caveat regarding the gas mass calculation is that, as well as assuming the same star formation law for the entire sample, the gas mass is estimated within 1~$R_e$. This introduces an aperture bias in the sense that galaxies with extended gas distributions will have their gas mass, and consequently their $M_b$, under-estimated \citep{Thomas+2004}. On the other hand, one of the criteria in \shade~sample selection requires the galaxies to have high \halpha flux (specifically, $f_\mathrm{H\alpha} > 5 \times 10^{-16} \mathrm{[erg \, s^{-1} \, cm^{-2} \, \text{\AA}^{-1} \, arcsec^{-2}]}$), This introduces a SFR bias that we have seen in Figure \ref{fig:SFR_Mbar_plot.a}. Using SFR to estimate the gas mass means the bias in SFR will lead to an over-estimation of the gas mass, and consequently an over-estimation of $M_{b}$. Without additional observations, it is difficult to quantify the combined effect of over-estimation of $M_b$ due to high \halpha~flux sample, and under-estimation of $M_b$ due to aperture bias. Therefore it will be important to avoid such estimation in future by pursuing a more accurate baryonic mass measurement through direct \HI\ observations of low-baryonic-mass galaxies to fully test the linearity of the scaling relation.

\begin{figure}
    \centering
    \includegraphics[scale=0.65]{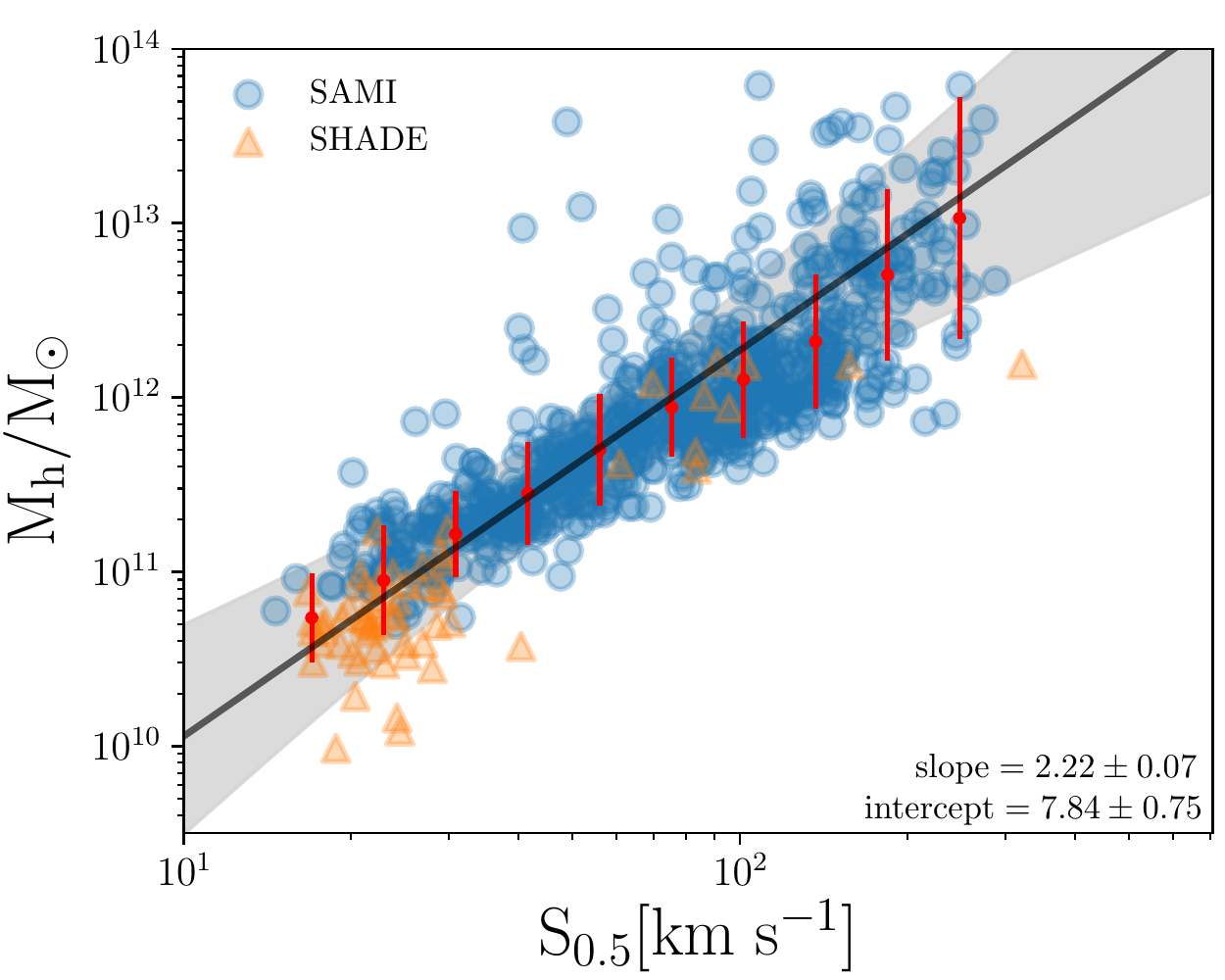}
    \caption{Halo mass version of the kinematic scaling relation. We estimate the halo mass of our sample by assuming a baryon-to-halo mass relation. The red points and error bars show the mean and standard deviation for the points in bins of $S_{0.5}$. We perform a linear regression (black solid line) to the binned values, with the shaded region representing the 3$\sigma$ range of the best-fit line. The slope and intercept of the inverse fit with $M_h$ as independent variable is given in the figure. We can see that, compared with the stellar mass and baryonic mass versions, although there is an increase of scatter at high masses ($10^{12}$--$10^{13}\,\rm{M_\odot}$), the curvature in the scaling relation is greatly reduced.} 
  \label{fig:HaloMass_ScalingRelation}
\end{figure}

\subsection{The effect of halo mass \texorpdfstring{\boldmath$M_h$}{Lg}}
\label{subsec:halomass}

In the standard CDM paradigm, the formation and kinematics of galaxies are under the influence of their dark matter halos. We cannot directly probe the dark matter independently of galaxy kinematics. However, we can use simple empirical baryon-to-halo mass relations from the literature, under reasonable assumptions, to estimate the dark matter halo masses ($M_h$). One such relation is given by \citet[][their Equation 2]{Moster+2010}, linking observed galaxy stellar mass with simulated halo mass using abundance matching. In that work, the stellar-to-halo mass relation is inferred for galaxies with masses $10^9<M_\star<10^{12}\,{\rm M_\odot}$. As we have seen, the baryonic mass in such high-mass galaxies tends to be star-dominated, while low-mass galaxies have higher gas content. Therefore, we assume that the stellar-to-halo mass relation in \citet{Moster+2010} is a good approximation of the baryonic-mass-to-halo mass relation, and that it can be applied to dwarf galaxies with $M_\star<10^{9}\,{\rm M_\odot}$ and so provide halo mass estimates for our whole sample\footnote{Fits from Table~2 of \citet{Moster+2010}, including the effect of scatter.}. 

Figure \ref{fig:HaloMass_ScalingRelation} shows the result of interpreting the stellar-to-halo mass ratio given by the relation of \citet{Moster+2010} as baryon-to-halo mass ratio and constructing the $\log M_{h}$--$ \log S_{0.5}$ scaling relation. We can see from the figure that, with this estimate of halo mass, the low-mass bend in the scaling relation has almost entirely disappeared. Moreover, the slight curvature of the  $\log M_{b}$--$ \log S_{0.5}$ relation at the high-mass end has also largely disappeared. Since we have made a number of assumptions, such as the applicability of the star formation law to early-type galaxies and the equivalence of the stellar-to-halo mass and baryon-to-halo mass relations, there is considerable systematic uncertainty in this figure and less constraint on systematic and sample biases in the resulting $\log M_{h}$--$ \log S_{0.5}$ relation than the original $\log M_{\star}$--$ \log S_{0.5}$ relation. To fit this relation, therefore, we perform a linear regression on the mean values in bins of $S_{0.5}$, shown as the red points in Figure \ref{fig:HaloMass_ScalingRelation} (although findings remain the same if we perform the fitting on the unbinned data). The fitting results are shown in the figure as well as Table \ref{table:fittingresult}. 

The fact that the $\log M_{h}$--$ \log S_{0.5}$ relation has the least amount of curvature over the range from giant to dwarf galaxies suggests that the estimated halo mass, rather than the stellar or baryonic mass, might be the most consistent quantity to use in a scaling relation that aims to unify galaxies of all morphologies. This suggests that the baryon-to-halo mass relation we have used conveniently captures the transition between gas-dominated, star-forming dwarf galaxies and dark-matter-dominated, quiescent massive galaxies. The reason why kinematics within one (optical) effective radius can successfully predict the halo mass remains to be explained, although it has been reported that total density profiles are almost isothermal \citep{cappellari+2015, poci+2017}. It is also worth noting that the $\log M_{h}$--$ \log S_{0.5}$ relation has a slope that is closest to the virial prediction $M \propto V_c^2$, in contrast to $\log M_{\star}$--$ \log S_{0.5}$ and $\log M_{b}$--$ \log S_{0.5}$ (see Table~\ref{table:fittingresult}). These findings underline the importance of dark matter halo mass in the construction of a unified galaxy kinematic scaling relation. It should be noted that while $M_h$ shows promising potential in rectifying the bend in the unified scaling relation, this result relies on quite a few significant assumptions. Direction \HI~observations are therefore essential for a proper test of these speculative conclusions.

Even though adding extra mass (both baryonic and dark matter) reduced the bend at low masses in the $\log M_{h}$--$ \log S_{0.5}$ relation, there is still a lower limit to the observed gas velocity dispersion. This limit occurs just at the mass range where the gas fraction in galaxies increases substantially. Understanding the driver of this lower limit to the gas velocity dispersion is important for understanding the formation and structure of low-mass dwarf galaxies.

\begin{figure}
    \centering
    \hspace*{-0.5cm}
    \includegraphics[scale=0.45]{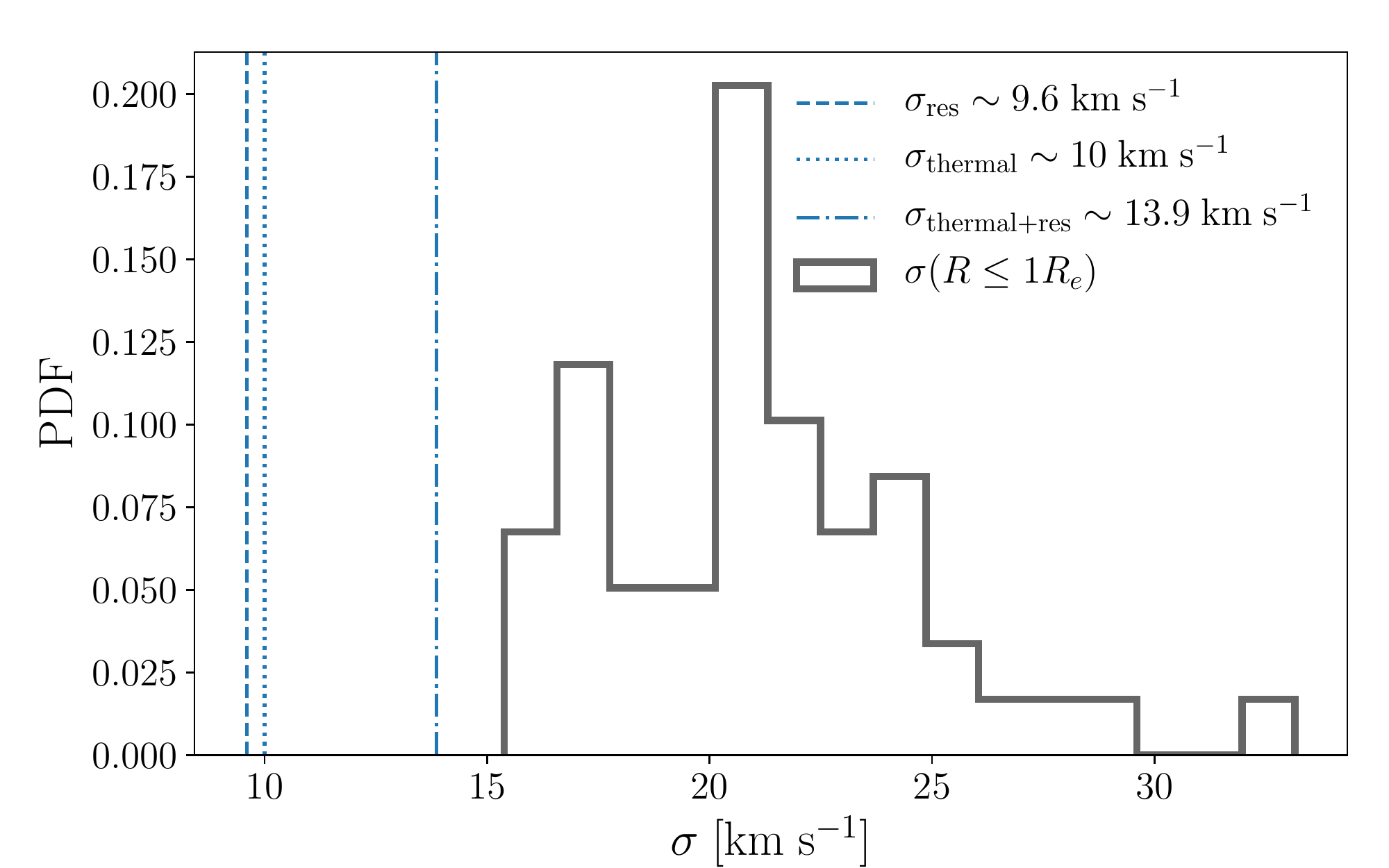}
    \caption{The velocity dispersion distribution for \shade galaxies with $\log{M_\star/{\rm M}_\odot} \lesssim 9.0$. The dashed line at $\sigma = 9.6$~\kms indicates the instrumental resolution ($\sigma_{\rm res}$). The dotted line at $\sigma = 10$~\kms represents the velocity dispersion contribution from the H{\sc ii} thermal contamination ($\sigma_{\rm thermal}$). The dot-dashed line at $\sigma = 13.9$~\kms represents the effective combined velocity dispersion limit $\sigma_{\rm thermal+res} = \sqrt{ \sigma^{2}_{\rm thermal} +  \sigma^{2}_{\rm res} }$.}
  \label{fig:Sigma_distribution}
\end{figure}

\subsection{The lower limits of \texorpdfstring{\boldmath$\sigma$}{Lg} and \texorpdfstring{\boldmath$S_{0.5}$}{Lg}}

As the bend in the $\log M_{\star}$--$ \log S_{0.5}$ scaling relation originates from the observed lower limit for $\sigma$ in the FJ relation, identifying the driver(s) behind $\sigma$ in low-mass dwarf galaxies is crucial to understanding the limitation of the $\log M_{\star}$--$ \log S_{0.5}$ scaling relation. 

Similar lower limits for $\sigma$ have been observed for \HI\ gas at around 10~\kms \citep{Ianjamasimanana+2012} and at around 20~\kms for \halpha gas \citep{Moiseev+2015}. Most recently, \citet{Varidel+2020} found that low-redshift star-forming galaxies from SAMI have  $14.1 < \sigma < 22.1$~\kms. Theoretical models also predict that the $\sigma$ of galaxies with low SFRs reach a minimum at $\sim$10~\kms \citep{Krumholz+2018}. 

It is clear from this study that below a stellar mass of $10^{8.6}$\,M$_{\odot}$ there is an excess of velocity dispersion relative to the gravitational potential from stellar mass. Figure~\ref{fig:Sigma_distribution} shows the distribution of the average velocity dispersion for \shade\ galaxies, excluding the SAMI control sample (i.e.\ for masses $M_\star \lesssim 10^9 {\rm M}_{\odot}$). This distribution has a mean velocity dispersion around 21.1~\kms and a 68\% confidence interval of [17.1,26.7]~\kms. This range is consistent with that from \citet{Varidel+2020}, with the slight difference perhaps stemming from the beam-smearing correction that they perform.

There are several possible drivers for enhanced velocity dispersions in late-type galaxies, such as merger events \citep{glazebrook2013}, gravitational instability \citep{Bournaud+2010, Bournaud+2014, Goldbaum+2015}, star-formation rate feedback \citep{Lehnert+2009, Lehnert+2013, Green+2010, Green+2014, Yu+2019}, and thermal contamination from H{\sc ii} regions (which contributes a velocity dispersion of $\sim$10~\kms; \citealt{Krumholz+2018}). While \citet{Krumholz+Burkhart2016} put forward star-formation feedback and gravitational instability as two models of turbulence, these models are most applicable to galaxies with relatively higher velocity dispersions ($\sigma > 50$~\kms) and higher SFR ($\rm SFR \gtrsim 10~M_{\odot} yr^{-1}$). Additionally, applications of these models often involve adding extra velocity dispersion of $\sim$15~\kms~ \citep[as seen in][]{Yu+2019, Varidel+2020} to account for the thermal contamination of \HII~regions. Therefore without \HII~contribution, neither of the two models can explain the observed velocity dispersion in \shade. 

The \HII~contamination argument \citep{Menon+2020} is plausible in explaining the velocity dispersion. This is because observations of extragalactic \halpha dynamics will always overlap with \HII~regions. The velocity dispersion of \HII~regions vary between 10~\kms~to 40~\kms~depending on their size and temperature \citep[e.g. 30 Doradus][]{Chu+Kennicutt1994}. Therefore the observed \halpha spectrum within an aperture will be a sum of Gaussian profiles, with minimum width of $\sim$10~\kms~or more. This means the minimum observed velocity dispersion is partially limited by the thermal expansion of \HII~regions. 

Another possible driver for the turbulence is supernovae feedback. Most recently, \citet{Bacchini+2020} determined the energy produced by supernovae explosions alone is sufficient to provide enough energy to match the kinetic energy measured from \HI~and CO observations of near by star-forming galaxies. 
They argue that in comparison to supernovae feedback, \HII~expansion is of secondary importance in driving the turbulence, based on the finding by \citet{Walch+2012} that \HII~expansion driven by stellar ionisation radiation can only explain about 2--4 \kms of the turbulence. Without additional data, determining which of these factors apply remains speculative and will need to be investigated further in future studies using both more extensive \HI~observations, stellar kinematics data and more sophisticated dynamical modelling.

\section{Summary}\label{s.sum}

In this work, we present the \shade\ survey, a high spectral resolution \halpha integral field survey of 69 dwarf galaxies in the local universe. We describe the survey goals, target selection, and data reduction process. 

We investigate the  $\log M_{\star}$--$ \log S_{0.5}$ kinematic scaling relation using these dwarf galaxy observations. We find that there exists a lower limit at $S_{0.5}\approx22.4\,\rm km\,s^{-1}$, which corresponds to a stellar mass limit of $M_\star \approx 10^{8.6}\,\rm M_{\odot}$. Above this limit, the scaling relation has a slope of $2.58\,\pm\,0.02$ and an intercept of $5.16\,\pm\,0.05$. This lower limit originates from an apparent lower limit in the observed \halpha velocity dispersion at $\sim$20\kms. These results are consistent with previous studies of the scaling relation using only SAMI data without the additional \shade\ observations. They suggest a physical origin of the low-mass bend in the \halpha version of  $\log M_{\star}$--$ \log S_{0.5}$ scaling relation. Using baryonic mass (based on estimating the gas mass from SFR measurements) reduces the severity of the bend in the scaling relation. This is partially due to the fact that, for their stellar mass, dwarf galaxies have higher sSFR compared to more massive galaxies. With some additional assumptions, the quantity that gives the most linear scaling relation is the estimated halo mass of galaxies, $M_h$. The $\log M_{h}$--$ \log S_{0.5}$ scaling relation is free of any bend at the low-mass end, has reduced curvature over the whole mass range, and brings galaxies of all masses and morphologies onto the virial relation.



\bibliographystyle{mnras}
\bibliography{astrobib} 

\section*{Acknowledgements}
We would like to thank the referee for their useful and constructive feedback. We would also like to thank Christoph Federrath, Sharon Meidt, and Ken Freeman for their advice and suggestions on the analyses. This work is based on observations made with ESO telescopes at the La Silla Paranal Observatory under programme IDs 0101.B-0505(A), 0101.B-0505(B) and 0101.B-05050(C). Part of this research was supported by the Australian Research Council Centre of Excellence for All Sky Astrophysics in 3 Dimensions (ASTRO 3D), through project CE170100013. DB is also supported by an Australia Government Research Training Program Scholarship. FDE acknowledges funding through the H2020 ERC Consolidator Grant 683184. LC is the recipient of an Australian Research Council Future Fellowship (FT180100066) funded by the Australian Government. The SAMI Galaxy Survey is based on observations made at the Anglo-Australian Telescope. The Sydney-AAO Multi-object Integral field spectrograph (SAMI) was developed jointly by the University of Sydney and the Australian Astronomical Observatory. The SAMI input catalogue is based on data taken from the Sloan Digital Sky Survey, the GAMA Survey and the VST ATLAS Survey. The SAMI Galaxy Survey is funded by the Australian Research Council Centre of Excellence for All-sky Astrophysics (CAASTRO), through project number CE110001020, and other participating institutions. The SAMI Galaxy Survey website is http://sami-survey.org/. Funding for SDSS and SDSS-II has been provided by the Alfred P.\ Sloan Foundation, the Participating Institutions, the National Science Foundation, the U.S.\ Department of Energy, the National Aeronautics and Space Administration, the Japanese Monbukagakusho, the Max Planck Society, and the Higher Education Funding Council for England. The SDSS Web Site is http://www.sdss.org/. SDSS is managed by the Astrophysical Research Consortium for the Participating Institutions. The Participating Institutions are the American Museum of Natural History, Astrophysical Institute Potsdam, University of Basel, University of Cambridge, Case Western Reserve University, University of Chicago, Drexel University, Fermilab, the Institute for Advanced Study, the Japan Participation Group, Johns Hopkins University, the Joint Institute for Nuclear Astrophysics, the Kavli Institute for Particle Astrophysics and Cosmology, the Korean Scientist Group, the Chinese Academy of Sciences (LAMOST), Los Alamos National Laboratory, the Max-Planck-Institute for Astronomy (MPIA), the Max-Planck-Institute for Astrophysics (MPA), New Mexico State University, Ohio State University, University of Pittsburgh, University of Portsmouth, Princeton University, the United States Naval Observatory, and the University of Washington.

\section*{Data Availability}
The measured kinematic quantities from \shade  data and ancillary data obtained from the SDSS DR12 used for the scaling relations analyses are available in the article in Appendix \ref{sec:ap} as well as on its online supplementary materials.

\onecolumn
\appendix
\section{Appendix}
\label{sec:ap}


\setlength{\tabcolsep}{2pt}
\begin{longtable}{llllllllcccc}

\caption{Quantities used in this work for analyses. $R_e$ (column 4) is obtained from \texttt{mgefit} fitting using SDSS DR12 $i$-band photometry; $Z$ (column 5) and $\log \rm{SFR}$ (column 9) are from SDSS DR12 database; $\sigma,\;\rm{V_{rot}},\;S_{0.5}$ are their uncertainties (columns 6, 7, 8) are calculated in this work following recipe outlined in section \ref{subsec:kin_measure}; $\rm\log M_{\star}$ (column 10) is calculated using $i$-band magnitude and $g-i$ colour following \citet{taylor+2011}; $\rm \log M_{b}$ (column 11) and  $\rm \log M_h$ (column 12) are estimated following \citet{Kennicutt1998} and \citet{Moster+2010} and are outlined in sections \ref{subsec: baryonicscalingrelation} and \ref{subsec:halomass} respectively.
}
\label{tab:long} \\

\hline 
ID         & Ra        & Dec       & Re       & Z      & $\sigma$        & $\rm V_{rot}$    & $S_{0.5}$        & $\rm\log SFR $ & $\rm\log M_{\star}/M_{\odot}$ & $\rm \log M_{b}/M_{\odot}$ & $\rm \log M_h/M_{\odot}$ \\
           & {\degr}   & {\degr}   & [arcsec] &        & [\kms]          & [\kms]           & [\kms]           &  [$\rm M_{\odot} yr^{-1}$]                                 &                               &                            &                          \\
(1)        & (2)       & (3)       & (4)      & (5)    & (6)             & (7)              & (8)              & (9)                               & (10)                          & (11)                       & (12)                     \\

\hline 
\endfirsthead

\multicolumn{10}{c}%
{{\bfseries \tablename\ \thetable{} -- continued from previous page}} \\
\hline 
ID         & Ra        & Dec       & Re       & Z      & $\sigma$        & $\rm V_{rot}$    & $S_{0.5}$        & $\rm\log SFR $ & $\rm\log M_{\star}/M_{\odot}$ & $\rm \log M_{b}/M_{\odot}$ & $\rm \log M_h/M_{\odot}$ \\
           & {\degr}   & {\degr}   & [arcsec] &        & [\kms]          & [\kms]           & [\kms]           & [$\rm M_{\odot} yr^{-1}$]                                  &                               &                            &                          \\
(1)        & (2)       & (3)       & (4)      & (5)    & (6)             & (7)              & (8)              & (9)                               & (10)                          & (11)                       & (12)                     \\

\hline 
\endhead

\hline \multicolumn{10}{l}{{Continued on next page}} \\ \hline
\endfoot

\hline \hline
\endlastfoot
                   
1   & 218.72581   & -0.34265   & 3.92   & 0.0060   & 15.4 $\pm$ 0.4   & 16.4 $\pm$ 1.2   & 19.3 $\pm$ 0.6   & -1.9   & 7.4   & 8.2   & 10.7     \\
2   & 235.93381   & -0.2383   & 2.96   & 0.0139   & 19.2 $\pm$ 0.8   & 12.7 $\pm$ 0.9   & 21.2 $\pm$ 0.8   & -1.3   & 7.6   & 8.6   & 10.9     \\
9   & 242.16249   & +0.45622   & 2.67   & 0.0246   & 21.3 $\pm$ 0.4   & 18.7 $\pm$ 3.4   & 25.1 $\pm$ 1.4   & -0.8   & 7.9   & 8.6   & 10.9     \\
42   & 187.41299   & +3.61269   & 6.70   & 0.0043   & 20.3 $\pm$ 0.3   & 14.0 $\pm$ 1.3   & 22.6 $\pm$ 0.4   & -1.7   & 7.9   & 8.7   & 11.0     \\
46   & 187.06652   & +1.82893   & 5.95   & 0.0030   & 16.2 $\pm$ 0.4   & 10.5 $\pm$ 2.9   & 17.8 $\pm$ 1.1   & -2.2   & 7.2   & 8.1   & 10.7     \\
55   & 231.20884   & +3.08143   & 2.57   & 0.0059   & 16.8 $\pm$ 0.3   & 7.1 $\pm$ 0.6   & 17.6 $\pm$ 0.3   & -1.8   & 7.3   & 7.9   & 10.6     \\
56   & 326.21562   & -8.76023   & 6.88   & 0.0043   & 18.2 $\pm$ 1.5   & 15.9 $\pm$ 3.8   & 21.4 $\pm$ 2.2   & -3.1   & 7.8   & 8.1   & 10.6     \\
59   & 7.93855   & -11.11585   & 1.23   & 0.0137   & 22.9 $\pm$ 1.0   & 2.5 $\pm$ 0.9   & 23.0 $\pm$ 1.0   & -1.5   & 7.3   & 7.7   & 10.5     \\
64   & 19.80951   & -9.59617   & 3.57   & 0.0064   & 16.9 $\pm$ 0.3   & 2.8 $\pm$ 0.8   & 17.1 $\pm$ 0.3   & -2.1   & 6.3   & 7.7   & 10.5     \\
128   & 202.94517   & -2.49656   & 5.80   & 0.0106   & 20.4 $\pm$ 0.3   & 30.0 $\pm$ 1.6   & 29.5 $\pm$ 0.8   & -1.1   & 8.1   & 8.9   & 11.1     \\
132   & 216.76992   & -1.72966   & 3.47   & 0.0060   & 19.0 $\pm$ 0.3   & 3.7 $\pm$ 0.4   & 19.1 $\pm$ 0.3   & -1.9   & 6.8   & 7.9   & 10.6     \\
136   & 234.77108   & +3.63822   & 2.30   & 0.0127   & 25.2 $\pm$ 0.6   & 20.0 $\pm$ 2.3   & 28.9 $\pm$ 1.0   & -1.1   & 8.0   & 8.6   & 10.9     \\
137   & 325.85917   & +0.60699   & 0.95   & 0.0269   & 20.4 $\pm$ 0.5   & 8.0 $\pm$ 1.9   & 21.2 $\pm$ 0.5   & -1.0   & 7.9   & 8.2   & 10.7     \\
148   & 314.53175   & -0.75542   & 1.47   & 0.0180   & 21.3 $\pm$ 0.3   & 10.1 $\pm$ 1.4   & 22.5 $\pm$ 0.4   & -1.3   & 7.8   & 8.2   & 10.7     \\
151   & 311.5677   & +0.08875   & 2.59   & 0.0120   & 17.3 $\pm$ 0.3   & 19.5 $\pm$ 3.1   & 22.2 $\pm$ 1.4   & -1.4   & 8.2   & 8.4   & 10.8     \\
152   & 311.84758   & +0.61849   & 1.83   & 0.0273   & 20.8 $\pm$ 1.5   & 6.4 $\pm$ 1.2   & 21.3 $\pm$ 1.5   & -1.3   & 7.5   & 8.2   & 10.7     \\
165   & 39.11989   & -0.97494   & 2.29   & 0.0083   & 26.4 $\pm$ 1.0   & 7.8 $\pm$ 1.3   & 26.9 $\pm$ 1.0   & -1.6   & 7.0   & 7.9   & 10.6     \\
171   & 165.64056   & +7.13526   & 2.14   & 0.0108   & 17.1 $\pm$ 0.3   & 6.8 $\pm$ 0.9   & 17.8 $\pm$ 0.3   & -1.5   & 7.5   & 8.1   & 10.7     \\
194   & 188.63253   & +13.50367   & 11.33   & 0.0027   & 16.6 $\pm$ 1.7   & 29.9 $\pm$ 0.9   & 26.9 $\pm$ 1.3   & -2.4   & 7.0   & 8.6   & 10.9 \\
200   & 169.86708   & +9.59564   & 3.34   & 0.0033   & 24.8 $\pm$ 1.2   & 18.2 $\pm$ 2.7   & 28.0 $\pm$ 1.6   & -2.1   & 6.4   & 7.6   & 10.4     \\
218   & 184.36686   & +12.93234   & 3.65   & 0.0069   & 16.5 $\pm$ 0.3   & 5.7 $\pm$ 1.5   & 17.0 $\pm$ 0.4   & -1.6   & 7.4   & 8.2   & 10.7     \\
231   & 233.68935   & +8.15691   & 1.39   & 0.0111   & 27.6 $\pm$ 0.6   & 10.3 $\pm$ 1.6   & 28.6 $\pm$ 0.8   & -1.3   & 7.8   & 8.1   & 10.7     \\
260   & 185.65594   & +8.29446   & 2.15   & 0.0046   & 17.7 $\pm$ 0.4   & 9.0 $\pm$ 0.7   & 18.8 $\pm$ 0.4   & -3.3   & 5.9   & 6.7   & 10.0     \\
271   & 190.74537   & +10.22723   & 1.47   & 0.0250   & 23.5 $\pm$ 0.8   & 24.9 $\pm$ 4.1   & 29.3 $\pm$ 1.1   & -1.1   & 8.3   & 8.5   & 10.9     \\
283   & 241.67081   & +6.58089   & 7.27   & 0.0058   & 20.9 $\pm$ 0.4   & 10.9 $\pm$ 0.7   & 22.2 $\pm$ 0.4   & -1.1   & 8.2   & 9.2   & 11.2     \\
284   & 220.11477   & +5.53190   & 2.94   & 0.0052   & 19.7 $\pm$ 0.4   & 5.9 $\pm$ 0.5   & 20.1 $\pm$ 0.3   & -1.8   & 5.8   & 7.8   & 10.5     \\
286   & 236.6024   & +4.48905   & 2.70   & 0.0062   & 20.2 $\pm$ 0.6   & 5.8 $\pm$ 0.8   & 20.7 $\pm$ 0.6   & -2.2   & 7.2   & 7.7   & 10.5     \\
288   & 234.26492   & +6.04018   & 2.41   & 0.0088   & 20.8 $\pm$ 0.6   & 9.3 $\pm$ 1.3   & 21.8 $\pm$ 0.7   & -1.9   & 7.7   & 8.1   & 10.7     \\
311   & 249.32163   & +17.40846   & 1.71   & 0.0249   & 23.5 $\pm$ 0.4   & 5.3 $\pm$ 1.4   & 23.8 $\pm$ 0.4   & -0.7   & 8.2   & 8.8   & 11.0     \\
314   & 348.37409   & -1.14795   & 1.95   & 0.0269   & 21.6 $\pm$ 0.4   & 8.2 $\pm$ 1.5   & 22.4 $\pm$ 0.5   & -1.0   & 7.5   & 8.5   & 10.8     \\
315   & 352.05124   & -1.06284   & 2.03   & 0.0087   & 24.6 $\pm$ 0.6   & 5.8 $\pm$ 1.1   & 25.0 $\pm$ 0.6   & -1.4   & 6.7   & 7.9   & 10.6     \\
319   & 312.28765   & -0.44892   & 1.56   & 0.0237   & 17.8 $\pm$ 0.2   & 3.5 $\pm$ 1.0   & 18.0 $\pm$ 0.2   & -1.0   & 7.5   & 8.2   & 10.7     \\
320   & 319.42655   & -0.30805   & 1.15   & 0.0195   & 21.6 $\pm$ 0.8   & 7.2 $\pm$ 2.0   & 22.2 $\pm$ 0.9   & -1.3   & 7.5   & 7.9   & 10.6     \\
322   & 315.69953   & +0.15969   & 1.88   & 0.0138   & 18.9 $\pm$ 0.4   & 8.3 $\pm$ 1.2   & 19.8 $\pm$ 0.5   & -1.0   & 7.0   & 8.4   & 10.8     \\
323   & 322.23272   & +0.05713   & 2.38   & 0.0309   & 25.6 $\pm$ 1.0   & 11.7 $\pm$ 2.5   & 26.9 $\pm$ 1.3   & -0.9   & 7.6   & 8.8   & 11.0     \\
327   & 336.33232   & -0.19058   & 1.27   & 0.0376   & 23.9 $\pm$ 2.3   & 4.3 $\pm$ 1.2   & 24.1 $\pm$ 2.3   & -1.3   & 7.9   & 8.3   & 10.7     \\
330   & 341.91433   & +0.32113   & 2.16   & 0.0157   & 21.2 $\pm$ 0.5   & 16.5 $\pm$ 5.7   & 24.2 $\pm$ 2.0   & -0.8   & 8.0   & 8.4   & 10.8     \\
331   & 311.70001   & +0.72771   & 1.07   & 0.0273   & 21.6 $\pm$ 0.3   & 2.6 $\pm$ 0.7   & 21.7 $\pm$ 0.3   & -1.2   & 7.9   & 8.2   & 10.7     \\
343   & 14.30249   & -0.36602   & 1.74   & 0.0094   & 20.2 $\pm$ 1.6   & 3.4 $\pm$ 0.9   & 20.3 $\pm$ 1.6   & -2.4   & 6.4   & 7.3   & 10.3     \\
344   & 359.10242   & +0.08913   & 1.97   & 0.0237   & 29.4 $\pm$ 2.3   & 9.8 $\pm$ 3.2   & 30.2 $\pm$ 2.7   & -1.7   & 7.1   & 8.2   & 10.7     \\
352   & 21.02392   & +0.98472   & 2.17   & 0.0069   & 24.0 $\pm$ 0.5   & 3.8 $\pm$ 0.6   & 24.2 $\pm$ 0.5   & -3.0   & 6.7   & 7.1   & 10.2     \\
424   & 206.4635   & +15.84032   & 3.01   & 0.0644   & 59.0 $\pm$ 2.3   & 52.3 $\pm$ 8.3   & 69.7 $\pm$ 3.1   & 1.3   & 10.1   & 10.5   & 12.1     \\
430   & 209.42121   & +15.37313   & 7.39   & 0.0042   & 16.0 $\pm$ 0.3   & 7.0 $\pm$ 1.0   & 16.7 $\pm$ 0.4   & -1.9   & 6.6   & 8.5   & 10.9     \\
431   & 231.48368   & +11.41037   & 2.91   & 0.0063   & 24.8 $\pm$ 3.5   & 5.9 $\pm$ 4.2   & 25.2 $\pm$ 4.0   & -2.1   & 7.0   & 7.8   & 10.5     \\
433   & 215.23256   & +15.35268   & 1.50   & 0.0182   & 20.6 $\pm$ 0.3   & 6.0 $\pm$ 1.7   & 21.0 $\pm$ 0.4   & -1.0   & 7.6   & 8.2   & 10.7     \\
435   & 234.54535   & +14.02722   & 1.84   & 0.0126   & 16.9 $\pm$ 0.4   & 3.4 $\pm$ 0.9   & 17.1 $\pm$ 0.4   & -1.6   & 7.8   & 8.1   & 10.7     \\
469   & 198.98461   & +17.76055   & 5.46   & 0.0032   & 21.1 $\pm$ 0.8   & 15.1 $\pm$ 1.2   & 23.7 $\pm$ 0.9   & -2.0   & 7.5   & 8.3   & 10.8     \\
496   & 190.99042   & +5.21342   & 3.53   & 0.0014   & 22.2 $\pm$ 3.4   & 14.7 $\pm$ 3.7   & 24.5 $\pm$ 3.7   & -3.2   & 5.7   & 6.9   & 10.1     \\
520   & 321.02057   & +0.98778   & 1.04   & 0.0196   & 33.2 $\pm$ 3.8   & 32.6 $\pm$ 9.1   & 40.4 $\pm$ 3.7   & -1.3   & 7.8   & 7.9   & 10.6     \\
106049   & 214.59013   & +0.94959   & 2.25   & 0.0261   & 23.0 $\pm$ 0.4   & 24.9 $\pm$ 2.9   & 28.9 $\pm$ 1.2   & -0.9   & 8.8   & 9.0   & 11.1     \\
296934   & 214.04425   & +1.54141   & 2.50   & 0.0530   & 36.7 $\pm$ 0.8   & 117.9 $\pm$ 10.0   & 91.1 $\pm$ 6.3   & 1.0   & 10.2   & 10.7   & 12.2     \\
319150   & 213.62262   & +1.81263   & 1.54   & 0.0254   & 20.7 $\pm$ 0.7   & 2.8 $\pm$ 0.5   & 20.8 $\pm$ 0.6   & -1.1   & 8.6   & 8.7   & 11.0     \\
511921   & 216.6746   & -1.14927   & 1.35   & 0.0307   & 26.2 $\pm$ 0.5   & 20.0 $\pm$ 3.1   & 29.8 $\pm$ 1.2   & -0.6   & 9.2   & 9.3   & 11.2     \\
594906   & 222.36208   & -0.1642   & 1.56   & 0.0410   & 62.2 $\pm$ 2.4   & 79.0 $\pm$ 11.4   & 83.6 $\pm$ 4.7   & 0.1   & 9.8   & 9.9   & 11.6     \\
9008500333   & 10.66479   & -9.51493   & 1.98   & 0.0575   & 96.3 $\pm$ 9.4   & 433.9 $\pm$ 61.3   & 321.6 $\pm$ 39.5   & -0.7   & 10.6   & 10.6   & 12.2     \\
9008500356   & 10.52508   & -9.60176   & 2.70   & 0.0529   & 69.1 $\pm$ 8.6   & 93.3 $\pm$ 15.8   & 95.6 $\pm$ 7.1   & -1.2   & 10.4   & 10.4   & 11.9     \\
9011900128   & 14.11509   & -1.38755   & 2.25   & 0.0425   & 32.8 $\pm$ 2.3   & 44.5 $\pm$ 4.3   & 45.4 $\pm$ 2.7   & --   & 9.5   & --   & --       \\
9016800065   & 18.83758   & +0.34296   & 2.56   & 0.0495   & 21.8 $\pm$ 1.8   & 113.7 $\pm$ 6.3   & 83.3 $\pm$ 4.2   & -0.2   & 9.8   & 10.1   & 11.7     \\
9016800314   & 18.59034   & +0.68980   & 2.65   & 0.0447   & 46.3 $\pm$ 12.2   & 56.0 $\pm$ 9.9   & 60.9 $\pm$ 11.8   & -0.8   & 9.9   & 9.9   & 11.6     \\
9091700123   & 355.19489   & -29.18565   & 1.99   & 0.0472   & 24.1 $\pm$ 0.9   & 42.3 $\pm$ 3.7   & 38.4 $\pm$ 1.7   & --   & 9.7   & --   & --       \\
9091700137   & 355.17484   & -29.27526   & 3.14   & 0.0514   & 94.0 $\pm$ 17.9   & 106.3 $\pm$ 13.3   & 120.4 $\pm$ 17.7   & --   & 10.7   & --   & --       \\
9091700444   & 354.86758   & -29.33298   & 2.43   & 0.0503   & 57.1 $\pm$ 3.3   & 47.4 $\pm$ 6.7   & 66.2 $\pm$ 3.1   & --   & 10.4   & --   & --       \\
9239900178   & 329.38877   & -7.59081   & 2.17   & 0.0565   & 93.5 $\pm$ 11.4   & 60.1 $\pm$ 9.7   & 102.7 $\pm$ 11.2   & -1.4   & 10.6   & 10.6   & 12.2     \\
9239900182   & 329.21017   & -7.68647   & 2.34   & 0.0552   & 112.7 $\pm$ 7.6   & 155.4 $\pm$ 20.5   & 157.4 $\pm$ 9.1   & 0.3   & 10.6   & 10.6   & 12.2     \\
9239900237   & 329.17712   & -7.66211   & 2.54   & 0.0554   & 46.9 $\pm$ 2.6   & 102.2 $\pm$ 6.5   & 86.2 $\pm$ 3.6   & 0.4   & 10.3   & 10.5   & 12.0     \\
9239900246   & 329.13968   & -7.7344   & 2.77   & 0.0604   & 25.4 $\pm$ 4.7   & 13.9 $\pm$ 4.0   & 27.2 $\pm$ 4.3   & --   & 10.4   & --   & --       \\
9239900370   & 329.39686   & -7.46727   & 2.09   & 0.0569   & 48.0 $\pm$ 9.8   & 26.6 $\pm$ 10.6   & 51.6 $\pm$ 9.6   & --   & 10.0   & --   & --       \\
9388000269   & 336.75073   & -30.89091   & 1.97   & 0.0568   & 78.7 $\pm$ 7.5   & 193.0 $\pm$ 10.7   & 157.5 $\pm$ 6.0   & --   & 10.5   & --   & --                                   

\end{longtable}


\bsp	
\label{lastpage}
\end{document}